\documentclass[11pt,reqno]{amsart}

\usepackage{packages}
\usepackage{notations}

\makeindex

\begin{document}

\title{From Painlevé to Zakharov-Shabat and beyond: \\
Fredholm determinants and integro-differential hierarchies}

\author{Alexandre Krajenbrink}

\address[Alexandre Krajenbrink]{  SISSA and INFN, via Bonomea 265, 34136 Trieste, Italy.   }
\email{\href{mailto:alexandre.krajenbrink@sissa.it}{alexandre.krajenbrink@sissa.it}}

\begin{abstract} As Fredholm determinants are more and more frequent in the context of stochastic integrability, we unveil the existence of a common framework in many integrable systems where they appear. This consists in a quasi-universal hierarchy of equations, partly unifying an integro-differential generalization of the Painlevé II hierarchy, the finite-time solutions of the Kardar-Parisi-Zhang equation, multi-critical fermions at finite temperature and a notable solution to the Zakharov-Shabat system associated to the largest real eigenvalue in the real Ginibre ensemble. As a byproduct, we obtain the explicit unique solution to the inverse scattering transform of the Zakharov-Shabat system in terms of a Fredholm determinant.
\end{abstract}

\maketitle

{\hypersetup{linkcolor=black}
\setcounter{tocdepth}{1}
\makeatletter
\def\l@subsection{\@tocline{2}{0pt}{2.5pc}{2.5pc}{}}
\makeatother
\tableofcontents
}

\section{Introduction}
\label{sec:intro}

One ambition of statistical mechanics consists in finding the universal underlying properties of physical systems. The ideal territories to achieve that are the so-called integrable systems which exhibit an extensive amount of conservation laws and where exact solutions are occasionally available therefore allowing to probe the exact physics of the problem \cite{babelon2003introduction}. Although advanced and poweful techniques such as the inverse scattering transform \cite{faddeev2007hamiltonian} and the Bethe ansatz \cite{gaudin2014bethe} have been developed, some territories still remain uncharted.  Among the various intriguing existing models, the determinantal point processes have attracted a growing attention during the last decades. In a nutshell, determinantal processes \cite{soshnikov2000determinantal} are the landmark of exclusion models \cite{spohn2012large} and they are sometimes referred to in the physics literature as Fermionic processes due to the Pauli exclusion principle. Mathematically, such particle models represent sets of random points so that every correlation function is given by the determinant of a certain kernel. \\

Remarkably, determinantal processes have been connected to a vast variety of fields both in mathematics and physics. In mathematics, they have been related to random matrix theory where the random eigenvalues form a determinantal process \cite{mehta2004random, anderson2010introduction, potters2019first}, to combinatorics \cite{baik2016combinatorics}, to Painlevé transcendents which are solutions to nonlinear differential equations \cite{tracy1994level, tracy1996orthogonal, borodin2002fredholm},  to integrable systems through the notion of $\tau$ function as defined by Jimbo, Miwa and Ueno \cite{jimbo1981monodromy}, to the Dyson Brownian motion \cite{dyson1962brownian}, to Riemann-Hilbert problems \cite{deift1999orthogonal, deift1997riemann}, to inverse scattering problems \cite{dyson1976fredholm, faddeev1976inverse} and many more. In physics, determinantal processes are intrinsically entangled with many theories such as free fermionic theory \cite{dean2015finite, dean2016noninteracting, le2018multicritical, stephan2019free}, quantum gravity \cite{fokas1992isomonodromy, fokas1991discrete, forrester2011non, di19932d, stanford2019jt}, directed polymers \cite{baik2000limiting, baik2018pfaffian}  and growth models \cite{imamura2004fluctuations, prahofer2000universal}, the Kardar-Parisi-Zhang equation \cite{sasamoto2010one, CalabreseDR, calabrese2011exact, ACQ, dotsenko, barraquand2020half}  and universality class \cite{quastel2015one}, soliton theory \cite{novikov1984theory} for nonlinear wave models such as the Kadomtsev-Petviashvili \cite{poppe1988fredholm} and the Korteweg-de Vries \cite{poppe1984fredholm} equations. More recently, an interest around determinantal processes has also grown in the field of statistics and machine learning \cite{kulesza2012determinantal} in particular for sampling purposes.\\

For all the problems where determinantal processes did arise, an important number of results were obtained for particular purposes related to combinatorics, functional analysis, integrability, probability or exact solvability of models. Nonetheless, it occurs that a few results have remained in the field where they were introduced. It is therefore natural to propose to gather and expand them to unveil more general frameworks in the hope to shed some new light on old or unresolved problems. This is the direction we undertake in this work where our ambition is to reveal that analogous results were obtained in seemingly unrelated situations: inverse scattering problems, the resolution of stochastic and deterministic nonlinear differential equations, random matrix theory and non-interacting systems in quantum mechanics. Our endeavor will be focused on a precise quantity: the Fredholm determinant arising in determinantal point processes.

\subsection*{Outline}
In Section~\ref{sec:lebesgueM}, we establish a common framework for Fredholm determinants generally related to the largest eigenvalue of random matrices or the rightmost fermion in a non-interacting system. This framework allows to generalize the classical analysis of Tracy-Widom and Brézin-Hikami interpreting notable Fredholm determinants as the $\tau$ function of non-linear systems and to introduce a quasi-universal hierarchy of functions. Our analysis unveils the existence of an infinite amount of conserved quantities in these systems highlighting their integrable structure. We also clarify some common results on matrix models relating their unitary, orthogonal and symplectic versions. 

In Section~\ref{sec:sigmaM}, we explore a more general structure of Fredholm determinants appearing in the study of linear statistics of random matrices, inhomogeneous full-counting statistics of free fermions and also finite-time solutions to the Kardar-Parisi-Zhang equation. Inspired by the work of Amir-Corwin-Quastel on an integro-differential generalization of the Painlevé II equation, we extend our framework to relate these inhomogeneous Fredholm determinants to integro-differential hierarchies and systems and show that they exhibit a richer structure of conserved quantities. 

In Section~\ref{sec:MultiFermion}, we apply our results to multi-critical fermions at finite temperature. This setting leads to a natural extension of the Painlevé II hierarchy whose definition and properties are recalled in Appendix~\ref{app:PIIhierarchy}. 
Our extended hierarchy is integro-differential and can be thus interpreted as non-local. The first member of the hierarchy was obtained by Amir, Corwin and Quastel, we explicitly compute its second member. Our construction naturally unveils some properties of the classical Painlevé II hierarchy.

In Section~\ref{sec:ZS}, we review and extend the recent results relating the distribution of the largest real eigenvalue of the real Ginibre ensemble and the Riemann-Hilbert problem associated to the celebrated Zakharov-Shabat system. In particular, we explicitly solve the Hilbert boundary value problem in terms of the hierarchy of functions established in Section~\ref{sec:lebesgueM} and show a one to one correspondence between a family of Fredholm determinants and the explicit unique solution of the Zakharov-Shabat system.

In Appendix~\ref{app:overview}, we present an overview of explicit examples where the Fredholm determinants occur and fit the framework of this work. It includes the linear statistics of Gaussian unitary, orthogonal and symplectic ensembles, 
 the eigenvalue statistics of the elliptic Ginibre ensembles and finally a few exact solutions of the Kardar-Parisi-Zhang equation.

\subsection*{Acknowledgments}  AK thanks P. Le Doussal and G.~Barraquand for enlightening discussions and ongoing collaborations and acknowledges support from ERC under Consolidator grant number 771536 (NEMO). AK thanks T. Gautié for carefully providing feedback and very useful comments on the manuscript.

\section{Fredholm determinants and counting statistics}\label{sec:lebesgueM}

We consider the  operators $K_s: \mathbb{L}^2(\R_+) \to  \mathbb{L}^2(\R_+)$  and $A_s:  \mathbb{L}^2(\R_+) \to \mathbb{L}^2(\R_+)$ depending on a real parameter $s$ such that their kernels are related as
\begin{equation}
\label{eq:KernelStructureMul}
\begin{split}
A_s(x,y)&=A(x+y+s)\\
K_s(x,y)&=\int_0^{+\infty} \rmd r \, A(x+r+s) A(y+r+s)
\end{split}
\end{equation}
for some real-valued function $A$ infinitely differentiable and vanishing exponentially fast towards $+\infty$. More concisely, we will write $K_s=A_s^2$ and we notice that both operators are self-adjoint by construction. The core object on which we will focus is the following Fredholm determinant 
\begin{equation}
\label{eq:FredDet}
\Det(I-K_s)_{\mathbb{L}^2(\R_+)}
\end{equation}
which can be represented by the following series 
\begin{equation}
\Det(I-K_s)_{\mathbb{L}^2(\R_+)}=1+\sum_{n=1}^\infty \frac{(-1)^n}{n!}\prod_{i=1}^n \int_{\R_+}\rmd x_i \det \left[K(x_i,x_j)\right]_{i,j=1}^n.
\end{equation}
A rigorous construction of these determinants and their relation to a determinantal point process with kernel $K$ can be found in \cite{anderson2010introduction,forrester2010log}.  We call these types of Fredholm determinants unitary-like determinants and further impose an orthogonality condition on the function $A$
\begin{equation}
\int_\R \rmd r A(s+r)A(r+s')=\delta(s-s')
\end{equation}
which can be interpreted as $A_s$ being its own inverse when considered on $\R$, additionally it ensures that $A_s$ is a Hilbert-Schmidt operator. As we will see, this will ensure that the operator $K_s$  uniquely defines the determinantal point process with correlation function $\rho_\ell(x_1,\dots,x_\ell)=\det[K_s(x_i,x_j)]_{i,j=1}^\ell$ for any $\ell\geq 1$. Indeed, the work of Soshnikov, Ref.~\cite[Theorem 3]{soshnikov2000determinantal}, asserts that a Hermitian locally trace-class operator $K_s$ uniquely defines a determinantal point process if and only if it is positive definite and bounded from above by the identity operator.\\

\begin{enumerate}
\item By construction $K_s$ is Hermitian (since $A$ is real) and locally trace-class (since we chose $A$ continuous).
\item By the Cauchy-Binet-Andreief formula \cite{andreief1883note, forrester2018meet}, for any $\ell \in \N$,
\begin{equation}
\det[K_s(x_i,x_j)]_{i,j=1}^\ell =\frac{1}{\ell !}\int_{\R_+^{\otimes \ell}}\prod_{j=1}^\ell \rmd u_j\big( \det[A(x_i+u_k+s)]_{i,j=1}^\ell\big)^2\geq 0
\end{equation}
proving its positivity.
\item To obtain that $K_s$  is bounded by above by the identity, we show that $K_s$ satisfies the reproducing property on $\R$, i.e. $K_s=K_s^2$. For any $x_1,x_2,s \in \R$, we have
\begin{equation}
\begin{split}
\int_\R \rmd y &K_s(x_1,y)K_s(y,x_2)\\
&=\int_\R \rmd y \int_{\R_+}\rmd r_1 \int_{\R_+}\rmd r_2 A(x_1+r_1+s)A(r_1+y+s)A(y+r_2+s)A(r_2+x_2+s)\\
&=\int_{\R_+}\rmd r_1 \int_{\R_+}\rmd r_2 A(x_1+r_1+s)\delta(r_1-r_2)A(r_2+x_2+s)\\
&=K_s(x_1,x_2)
\end{split}
\end{equation}
implying that $K_s\leq 1$.
\end{enumerate}

For what follows, we will need the following invertibility result for $K_s$.
\begin{result}[Existence of the resolvent of $K_s$]\label{result:existenceResol}
The operator $I-K_s$ is invertible on  $\mathbb{L}^2(\R_+)$.
\end{result}
\begin{proof}
To obtain this, we will proceed by apagogy supposing there exists a non-zero $f$ on $\R_+$ such that $K_sf=f$. From the decomposition of $K_s$ in terms of $A_s$ and the orthogonality property of $A$, we have for all real positive $x$
\begin{equation}
\int_{\R_+} \rmd y\, K_s(x,y)f(y)=f(x)-\int_{\R_-}\rmd z \, \int_{\R_+} \rmd y \, A(x+z+s)A(z+y+s)f(y)
\end{equation}
Since $K_sf=f$, it yields that the last term is equal to 0. Multiplying this term by $f(x)$ and integrating over $x\in \R_+$, we obtain the $\mathbb{L}^2$-norm of $ \R_- \ni u \mapsto (A_s f)(u)$ is equal to 0. We thus obtain for all $u<0$
\begin{equation}
\label{eq:apagogy}
\int_{\R_+}\rmd x A(u+s+x)f(x)=0
\end{equation}
As from the standard arguments, e.g. \cite[Proof of Lemma~2.1]{baik2020largest} or \cite[Proof of Lemma~6.15]{baik2016combinatorics},  assuming $A$ to have analytic properties, Eq.~\eqref{eq:apagogy} should also hold for $u>0$ and therefore we have $f=0$: the contradiction.
\end{proof}

We will also need the following extension to operators we call thinned operators. These operators appear in the context of random matrix theory when considering thinned ensembles where all eigenvalues are independently removed with a probability $1-\gamma\in [0,1]$ and also in the context of quantum mechanics when evaluating the entanglement entropy in a free fermionic theory, see Refs.~\cite{calabrese2004entanglement, calabrese2009entanglement, calabrese2015random}.

\begin{result}[Generalization to thinned operators]
All above results hold upon replacing $K_s$ by $\gamma K_s$ with $\gamma \in [0,1]$. In particular $I-\gamma K_s$ is invertible on $\mathbb{L}^2(\R_+)$ for all $\gamma\in [0,1]$. This implies that $A_s$ is bounded by above by the identity and that $I\pm \sqrt{\gamma} A_s$ are invertible on $\mathbb{L}^2(\R_+)$ since $I-\gamma K_s=(I-\sqrt{\gamma}A_s)(I+\sqrt{\gamma}A_s)$.
\end{result}

The structure of the kernel $A_s$ provides a few calculation rules that we will extensively use throughout this work. For all $x,y,s\in \R$, since   $A_s(x,y)=A(x+y+s)$, we have 
\begin{equation}
(\partial_x-\partial_y)A_s(x,y)=0.
\end{equation}
Defining the differential operator $D$ on $\mathbb{L}^2(\R_+)$ and its transpose $D^\intercal$, this relation is shorthanded as $DA=AD^\intercal$.
Assuming all functions vanish sufficiently fast at $+\infty$, the integration by part on $\R_+$ reads $D^\intercal=-D-\ket{\delta}\bra{\delta}$ where $\ket{\delta}\bra{\delta}$ is the projector to  zero. Applied to the operator $A_s$, it yields the following rule
\begin{equation}
DA_s+A_sD=-A_s\ket{\delta}\bra{\delta} .
\end{equation}
A second rule coming from the structures of $K_s$ and $A_s$ is that the derivative of $K_s$ with respect to $s$ yields a rank-one operator
\begin{equation}
\partial_s K_s=-A_s\ket{\delta}\bra{\delta}A_s .
\end{equation}

Our main result for the family of Fredholm determinants of Eq.~\eqref{eq:FredDet} is the existence of a hierarchy of scalar-valued  functions.
\begin{definition}[Conjugated functions]
For any $s$ in $\R$ and $p$ in $\N$, we define the scalar-valued functions 
\begin{equation}
\label{eq:ConjVar}
q_p=\bra{\delta} \frac{A_s^{(p)}}{I-K_s}\ket{\delta}, \qquad u_p=\bra{\delta} A_s\frac{I}{I-K_s}A_s^{(p)}\ket{\delta},
\end{equation}
where $A_s^{(p)}$ stands for the $p$-th derivative of $A_s$ with respect to $s$ with the convention $A_s^{(0)}=A_s$. The definitions imply the following boundary conditions
\begin{equation}
q_p\sim A^{(p)}(s), \qquad u_p\sim \int_{\R_+}\rmd r \, A^{(p)}(r+s)A(r+s), \qquad s\to +\infty.
\end{equation}
since the kernel $K_s$ decays exponentially fast for large $s$ and thus the resolvent is at first order the identity. Besides, the maps $s\mapsto \{q_p,u_p\}_{p\in \N}$ are smooth due to the analyticity property of the resolvent and the differentiability of $A$.
\end{definition}
In terms of explicit integrals, Eqs.~\eqref{eq:ConjVar} are written as 
\begin{equation}
\label{eq:HomoDefInt}
\begin{cases}
q_p=\int_{\R_+} \rmd y \, A^{(p)}(s+y)(I-K_s)^{-1}(y,0)\\
u_p=\iint_{\R_+^2}\rmd y \, \rmd z\, A(s+y)(I-K_s)^{-1}(y,z)A^{(p)}(z+s)
\end{cases}
\end{equation}
These functions have appeared multiple times already in the literature, but in very specific context, see e.g. Refs.~\cite{tracy1994level, brezin1998level, le2018multicritical}. The question of the asymptotics of these functions for large negative $s$ is in general complicated and unsolved for many problems. As we will see subsequently, the set $\{q_p,u_p\}_{p\in \N}$ forms a hierarchy.\\

Our first important result is the relation between the Fredholm determinant \eqref{eq:FredDet} and the first members of the hierarchy \eqref{eq:ConjVar} which we call the $\tau$-representation of the Fredholm determinant.
\begin{result}[$\tau$-representation of the Fredholm determinant]
\label{res:tau1}
The following derivatives of the logarithm of the Fredholm determinant hold
\begin{equation}
\partial_s \log \Det(I-K_s)=u_0, \qquad \partial_s^2 \log \Det(I-K_s)=-q_0^2 .
\end{equation}
\end{result}

\begin{proof}
Using the identity for the derivative of the logarithm of a Determinant and the cyclicity of the trace
\begin{equation}
\begin{split}
\partial_s \log \Det(I-K_s)&=-\Tr(\frac{I}{I-K_s}\partial_s K_s)\\
&=\Tr(\frac{I}{I-K_s}A_s\ket{\delta}\bra{\delta}A_s)\\
&=\bra{\delta}A_s\frac{I}{I-K_s}A_s\ket{\delta}\\
&=u_0
\end{split}
\end{equation}
We defer the proof of the second identity for the next result. 
\end{proof}

\begin{remark}
The above Result \ref{res:tau1} has an additional interpretation in terms of a Poisson point process. From the asymptotics of the kernel $K_s$, we have upon integration
\begin{equation}
\Det(I-K_s)=\exp\big(-\int_{s}^{+\infty} u_0(t) \rmd t\big).
\end{equation}
For a determinantal point process with kernel $K_s$, the left hand side represents the emptiness probability of the interval $[s,+\infty)$ or equivalently the probability that the right-most point lies to the left of $s$. Hence, denoting $x_{{\rm DPP}, \max}$ the right-most point of the determinantal process we have $\PP(x_{{\rm DPP}, \max} \leqslant s)=\Det(I-K_s)$.

Now consider a Poisson point process on $\R$ with density $u_0(t)$. Since the random points are drawn independently, the distribution of the right-most point denoted $x_{{\rm Pois},\max}$ is given by, see e.g. Ref.~\cite[Example~2.5]{johansson2005random},
 $\PP(x_{{\rm Poi}, \max} \leqslant s)=\exp\big(-\int_{s}^{+\infty} u_0(t) \rmd t\big)$. Hence we obtain the equality between the extremal laws
\begin{equation}
\PP(x_{{\rm Poi}, \max} \leqslant s)=\PP(x_{{\rm DPP}, \max} \leqslant s).
\end{equation}

The two processes are nonetheless not equal since thinning the Poisson process with a probability $1-\gamma \in [0,1]$ induces the change $u_0 \to \gamma u_0$ and thinning the determinantal process induces a different change
\begin{equation}
u_0=\bra{\delta} \frac{K_s}{I-K_s}\ket{\delta} \to \bra{\delta} \frac{\gamma K_s}{I-\gamma K_s}\ket{\delta}  \neq \gamma u_0.
\end{equation}
\end{remark}
~\\
The second main result is the differential relation between the members of the hierarchy $\{q_p,u_p\}_{p\in \N}$.
\begin{result}[Hierarchy of equations]
For all $p$ in $\N$, the following infinite recursion holds
\begin{equation}
\label{eq:HomoQuasiUniversalHie}
q_p'=q_{p+1}-q_0 u_p, \qquad u_p'=-q_0 q_p .
\end{equation}
The prime $'$ stands for the derivative with respect to $s$.
\end{result}
\begin{proof}
The first identity is readily obtained by differentiating the product of two kernels and using the identity of the derivative of the resolvent $\partial_s (I-K_s)^{-1}=(I-K_s)^{-1} \partial_s K_s (I-K_s)^{-1}$.
\begin{equation}
\begin{split}
q_p'&=\bra{\delta} \frac{A_s^{(p+1)}}{I-K_s}\ket{\delta}-\bra{\delta} A_s^{(p)} \frac{I}{I-K_s}A_s\ket{\delta}\bra{\delta}A_s\frac{I}{I-K_s}\ket{\delta}\\
&=q_{p+1}-q_0 u_p
\end{split}
\end{equation}
The second identity is obtained by first using a single element decomposition
\begin{equation}
\begin{split}
u_p'&=\frac{1}{2}\bra{\delta} \partial_s \big(\frac{I}{I-A_s}-\frac{I}{I+A_s}\big) A_s^{(p)}\ket{\delta}+\bra{\delta}\frac{I}{I-K_s}DA_s^{(p)}\ket{\delta}.\\
\end{split}
\end{equation}
Differentiating with respect to $s$ and using the Ferrari-Spohn formula for both $A_s$ and $-A_s$, see Lemma~\ref{lemma:FerrariSpohn},  we get 
\begin{equation}
\begin{split}
u_p'&=-\bra{\delta}\frac{A_s}{I-K_s}\ket{\delta} \bra{\delta}\frac{A_s^{(p)}}{I-K_s} \ket{\delta}\\
&=-q_0q_p
\end{split}
\end{equation}
\end{proof}
This hierarchy can be viewed as a generalization of the one in Ref.~\cite[Eq.~(109) arXiv version]{doussal2018multicritical}. The structure of the hierarchy additionally implies the existence of an infinite number of quadratic conserved quantities given as follows. 
\begin{result}[Flow invariance]\label{result:FlowInvarianceI}
For all $n$ in $\N$, the following quadratic quantity is invariant within the hierarchy
\begin{equation}
\mathcal{I}_n=u_{2n+1}+\frac{1}{2}\sum_{k=0}^{2n} (-1)^{k+1}[u_k u_{2n-k}-q_k q_{2n-k}]=0
\end{equation}
\end{result}
\begin{proof}
We differentiate the identity and obtain a telescopic equation equating to zero. Integrating and using that $u_p, q_p$ vanish at $+\infty$, we obtain the flow invariance.
\begin{equation}
\begin{split}
\mathcal{I}_n'&=-q_0q_{2n+1}-\frac{1}{2}\sum_{k=0}^{2n} (-1)^{k+1}[q_kq_{2n-k+1}+q_{k+1}q_{2n-k}]\\
&=0
\end{split}
\end{equation}
\end{proof}
For $n=0$, this result in case of the Tracy-Widom distribution and its extensions already appeared in e.g. \cite{tracy1994level, imamura2004fluctuations}. For general $n$, this is again a generalization of the result of Ref.~{\cite[Eq.~(114) arXiv version]{doussal2018multicritical}}.

\subsection{Fredholm determinants extended with a rank-one perturbation}\label{subsec:OrthoSymplecLike}
The reason we named Fredholm determinants of the type \eqref{eq:FredDet} unitary-like is because they usually arise in problems related to Hermitian random matrices enjoying a unitary symmetry. Among the additional ensembles of interest in random matrix theory and affiliated problems are orthogonal and symplectic ensembles where the structure of the corresponding Fredholm determinants are closely related to their unitary counterpart.  In this Section, we will introduce three counterparts of the unitary determinants: the orthogonal, orthogonal-thinned and symplectic ones and show that they also exhibit a $\tau$-representation in terms of the hierarchy constructed. 
\subsubsection{Determinantal representation}

Just before introducing the three types of determinants, we will require two further results. The first one is the matrix determinant lemma which enables to calculate the Fredholm determinant of a given kernel with a rank-one perturbation and the second one a result from Ferrari and Spohn \cite{ferrari2005determinantal}. 
\begin{lemma}[Matrix determinant lemma]
\label{lemma:MatDetLem}
Given a kernel $K_s$ perturbed by a rank-one operator $\ket{f}\bra{g}$ of kernel $(x,y)\mapsto f(x)g(y)$, we have 
\begin{equation}
\Det(I-K_s-\ket{f}\bra{g})=\Det(I-K_s)\big(1-\bra{f}\frac{I}{I-K_s}\ket{g}\big)
\end{equation}
The integral representation of the inner product reads
\begin{equation}
\bra{f}\frac{I}{I-K_s}\ket{g}=\int_{\R_+^2}\rmd x  \rmd y \, f(x) (I-K_s)^{-1}(x,y)g(y)
\end{equation}
The extension where the kernel $K_s$ is perturbed by a sum of rank-one operators is given in the Appendix in Lemma~\ref{lemma:MatDetLemEx}.
\end{lemma}
Anticipating  the particular form of the inner products that will appear for the orthogonal and symplectic determinants, the following result will allow us further simplifications in the Fredholm determinants. 
\begin{lemma}[Ferrari-Spohn, {\cite{ferrari2005determinantal}}]
The following identity holds
\begin{equation}
\label{eq:FerSp}
b=\bra{1}\frac{1}{I+A_s}\ket{\delta}=\frac{\Det(I-A_s)}{\Det(I+A_s)}.
\end{equation}
A proof for this identity uses the exact same arguments as Refs.~\cite[Eq.~(19)]{ferrari2005determinantal} and \cite[Eq.~(6.8)]{baik2020largest}, the only required arguments are that the operator $A_s$ has a kernel of the type $A_s(x,y)=A_s(x+y)$ and the function $A$ has a sufficient decay at $+\infty$. By symmetry, the second equality also holds upon replacing $A_s \to -A_s$. The integral representation of \eqref{eq:FerSp} reads
\begin{equation}
b=\int_{0}^{+\infty} \rmd x \, (I+A_s)^{-1}(x,0).
\end{equation}
\end{lemma}

The three types of determinants which are interest are the following. We will report some of the standard calculation of each of the cases.

\begin{definition}[Orthogonal-like determinant]\label{def:OrthoLike}
The orthogonal-like determinant is a Fredholm determinant of an operator with kernel 
\begin{equation}
K_s^{(\rm ortho)}(x,y)=K_s(x,y)+A(s+x)\big(1-\int_{0}^{+\infty}\rmd r \, A(s+r+y)\big)
\end{equation}
In terms of bra-ket notations, the Fredholm determinants is expressed and simplified as
\begin{equation}
\begin{split}
\Det(I-K_s-&A_s\ket{\delta}\bra{1}(I-A_s))\\
&=\Det(I-K_s)\big(1-\bra{1}\frac{I-A_s}{I-K_s}A_s\ket{\delta}\big)\\
&= \Det(I-K_s)\bra{1}\frac{I}{I+A_s}\ket{\delta}\\
&=\Det(I-A_s)^2
\end{split}
\end{equation}
This type of orthogonal-like determinant has appeared for instance in Refs.~\cite{baik2020largest, ferrari2005determinantal, sasamoto2005spatial}.
\end{definition}
\begin{definition}[Symplectic-like determinant] \label{def:SymplLike}
The symplectic-like determinant is a Fredholm determinant of an operator with kernel
\begin{equation}
K_s^{(\rm sympl)}(x,y)=K_s(x,y)-\frac{1}{2} A(s+x)\int_{0}^{+\infty}\rmd r \, A(s+r+y)
\end{equation}
In terms of bra-ket notations, the Fredholm determinants is expressed and simplified as
\begin{equation}
\begin{split}
\Det(I-K_s+&\frac{1}{2} A_s\ket{\delta}\bra{1}A_s)\\
&=\Det(I-K_s)\big(1+\frac{1}{2}\bra{1}\frac{K_s}{I-K_s}\ket{\delta}\big)\\
&=\frac{1}{4}\Det(I-K_s)\big(2+\bra{1}\frac{I}{I+A_s}\ket{\delta}+\bra{1}\frac{I}{I-A_s}\ket{\delta}\big)\\
&=\frac{1}{4}\big(\Det(I+A_s)+\Det(I-A_s)\big)^2
\end{split}
\end{equation}
This type of symplectic-like determinant has appeared for instance in Refs.~\cite{gueudre2012directed, barraquand2020half}.
\end{definition}
\begin{definition}[Orthogonal-thinned-like determinant] \label{def:OrthoThinnedLike}
The orthogonal-thinned-like determinant is a Fredholm determinant of an operator with kernel
\begin{equation}
K_s^{(\rm ortho-thinned)}(x,y)=K_s(x,y)+A(s+x)\big(\sqrt{\alpha}-\int_{0}^{+\infty}\rmd r \, A(s+r+y)\big)
\end{equation}
for some constant $\alpha$. In terms of bra-ket notations, the Fredholm determinants is expressed and simplified as
\begin{equation}
\begin{split}
\Det(I-&K_s-A_s\ket{\delta}\bra{1}(\sqrt{\alpha}\,  I-A_s))\\
&=\Det(I-K_s)\big(1-\bra{1}\frac{\sqrt{\alpha}\,  I-A_s}{I-K_s}A_s\ket{\delta}\big)\\
&= \Det(I-K_s)\big(\frac{\sqrt{\alpha}+1}{2}\bra{1}\frac{I}{I+A_s}\ket{\delta}-\frac{\sqrt{\alpha}-1}{2}\bra{1}\frac{I}{I-A_s}\ket{\delta}\big)\\
&=\frac{\sqrt{\alpha}+1}{2}\Det(I-A_s)^2-\frac{\sqrt{\alpha}-1}{2}\Det(I+A_s)^2
\end{split}
\end{equation}
This type of orthogonal-thinned-like determinant has appeared for instance in Refs.~\cite{bohigas2009deformations, bothner2018large, baik2020largest}. The connexion with the thinned ensembles of random matrix theory comes upon the replacement of the operator $A_s$ by its thinned version $\sqrt{\gamma} A_s$ for $\gamma \in [0,1]$ and the identification $\alpha=\gamma$. Since our identity is more general, we choose to keep the free parameter $\alpha$. 
\end{definition}

\subsubsection{Pfaffian representation}
For completeness, we additionally present in this Section a representation of the symplectic and orthogonal determinants as Pfaffians of $2\times 2$ matrix-valued kernels. One of the motivation lies about the generalization of the determinantal point processes to Pfaffian point processes whose correlation functions are expressed in terms of a Pfaffian of a correlation kernel $K$, i.e. for all $k$,  $\rho_k(x_1,\dots,x_k)={\rm Pf} \left[K(x_i,x_j)\right]_{i,j=1}^k$. We recall the definition of the Pfaffian of an anti-symmetric matrix $A$ of size $2N \times 2N$ 
\begin{equation}
{\rm Pf}(A)=\sqrt{{\rm Det}(A)}=\sum_{\substack{\sigma\in S_{2N}\\ \sigma(2p-1)<\sigma(2p)}}{\rm sign}(\sigma)\prod_{p=1}^{N}A_{\sigma(2p-1),\sigma(2p)}
\end{equation}
Here our focus will be in the case where the kernel $K$ is not scalar valued but rather $2\times 2$ matrix-valued, and we represent $K$ with $2\times 2$ blocks as follows
\begin{equation}\label{eq:correctionJuly3}
K(x,y)=
 \left(\begin{array}{cc} 
K_{11}(x,y)&K_{12}(x,y) \cr
  K_{21}(x,y)  & K_{22}(x,y) \cr
\end{array}\right).
\end{equation}
For such a kernel to be anti-symmetric, we shall also require, $K_{11}(x,y)=-K_{11}(y,x)$, $K_{22}(x,y)=-K_{22}(y,x)$,  $K_{21}(x,y)=-K_{12}(y,x)$. With this definition, the correlation function $\rho_k(x_1,\dots,x_k)={\rm Pf} \left[K(x_i,x_j)\right]_{i,j=1}^k$ is well defined since the Pfaffian is evaluated for the $k\times k$ matrix containing anti-symmetric $2\times 2$ blocks so that the overall $2k\times 2k$ matrix is antisymmetric.\\

The direct generalization of the Fredholm determinant studied in this work is the Fredholm Pfaffian. For a kernel $K$, defining the matrix kernel $J(r,r')=\big(\begin{array}{cc}
0 & 1 \\ 
-1 & 0
\end{array} 
\big)\mathds{1}_{r=r'}$, the Fredholm Pfaffian of $K$ stands for the following series
\be
{\rm Pf}\left[J-K\right]_{\mathbb{L}^2(\Omega)}
=1+ \sum_{n_s=1}^\infty \frac{(-1)^{n_s}}{n_s!}  \prod_{p=1}^{n_s} \int_\Omega \rmd r_p \;  {\rm Pf}[K (r_i,r_j)]_{i,j=1}^{n_s}.
\ee
The relation between a Pfaffian and a Determinant extends to their Fredholm counterpart as
\begin{equation}
\mathrm{Pf}[J-K]^2=\mathrm{Det}(I+JK)
\end{equation}
see Ref.~\cite[Lemma~8.1]{rains2000correlation}. For the further properties of Fredholm Pfaffians we refer the reader to \cite[Section~8]{rains2000correlation},
as well as e.g. \cite[Section~2.2]{baik2018pfaffian}, \cite[Appendix~B]{ortmann2017pfaffian} and \cite[Appendix~G]{le2012kpz}.\\

As in standard in the random matrix literature, see \cite{tracy1996orthogonal, forrester2000painlev, a2005matrix}, we express the symplectic and orthogonal determinants as Fredholm Pfaffians and defer for readibility the standard derivation of these identities to Appendix~\ref{app:pfaffOperator}. 

\begin{result}[Symplectic-like Pfaffian]\label{result:SymplecPfaff}
Defining the antisymmetric operator $B^{(\rm sympl)}$ defined in terms of the kernel of Def.~\ref{def:SymplLike} such that
\begin{equation}
B^{(\rm sympl)}=\frac{1}{2}D^{-1} K_s^{(\rm sympl)}=\frac{1}{2}D^{-1}K_s+\frac{1}{4}A_s\ket{1}\bra{1}A_s
\end{equation}
then we have the equality between the Fredholm determinant of scalar-valued kernel and the Fredholm Pfaffian of matrix-valued kernel (we omit the subscript for readability)
\begin{equation}
\Det(I-K_s^{(\rm sympl)})={\rm Pf} \left(J-
\begin{bmatrix}
B &-BD^\intercal\\ 
-DB &  DBD^\intercal\end{bmatrix}\right)^2
\end{equation}
Another representation in terms of a Fredholm Pfaffian involving a $\delta'$ operator is additionally available in Proposition~\ref{prop:barraquand}.
\end{result}
\begin{result}[Orthogonal-like Pfaffian]\label{result:OrthoPfaff}
Defining the antisymmetric operator $B^{(\rm ortho)}$ defined in terms of the kernel of Def.~\ref{def:OrthoLike} such that
\begin{equation}
B^{(\rm ortho)}=D^{-1}K_s+\frac{1}{2}A_s\ket{1}\bra{1}A_s +\frac{1}{2}\big( \ket{1}\bra{1}A_s-A_s\ket{1}\bra{1}\big)
\end{equation}
then we have the equality between the Fredholm determinant of scalar-valued kernel and the Fredholm Pfaffian of matrix-valued kernel (we omit the subscript for readability)
\begin{equation}
\Det(I-K_s^{(\rm ortho)})={\rm Pf} \left(J-
\begin{bmatrix}
B-\varepsilon &-BD^\intercal\\ 
-DB &  DBD^\intercal\end{bmatrix}\right)^2
\end{equation}
where $\varepsilon$ is an antisymmetric operator with kernel
\begin{equation}
\varepsilon(x,y)=\varepsilon(x-y)=\frac{1}{2}\sgn(x-y)=\Theta(x-y)-\frac{1}{2}
\end{equation}
so that
$D\varepsilon=I$ is the identity, $\varepsilon\ket{\delta}_0=\frac{1}{2}\ket{1}$ and $\varepsilon\ket{\delta}_\infty=-\frac{1}{2}\ket{1}$.
\end{result}

\subsubsection{$\tau$-representation} Having introduced the determinantal and Pfaffian representations of the three families of determinants, we now relate them to the first members of the hierarchy of function. To do so, we first establish that the inner product \eqref{eq:FerSp} admits a representation in terms of $q_0$. 
\begin{result}[$\tau$-representation of $b$]
The following differential and integral representations of $b$, defined in Eq.~\eqref{eq:FerSp}, in terms of the first function of the hierarchy $q_0$ hold
\begin{equation}
\label{eq:FerrSpohnPainlevQ}
b'=q_0b, \qquad b=e^{-\int_s^{+\infty} \rmd r \, q_0(r)}
\end{equation}
where the prime stands for the derivative with respect to $s$.
\end{result}
\begin{proof}
The differential equation is a direct consequence of the Ferrari-Spohn derivative formula \ref{lemma:FerrariSpohn} using the operator identity $D\ket{1}=0$. The constant of integration is fixed by comparing asymptotics for large positive $s$ using the exponential decay of $A$ towards $+\infty$. 
\end{proof}

Equipped with the representation of the inner product in terms of $q_0$, we obtain the $\tau$-representation of the three families of determinants as follows. 
\begin{result}[$\tau$-representations of the perturbed kernels] The following Fredholm determinants admit a $\tau$-representation.\\
\begin{itemize}
\item Orthogonal-like determinant
\begin{equation}
\begin{split}
\Det(I-K_s-&A_s\ket{\delta}\bra{1}(I-A_s))\\
&=\exp\big(-\int_s^{+\infty} \rmd r \, [(r-s)q_0(r)^2+q_0(r)]\big)
\end{split}
\end{equation}
\item Symplectic-like determinant
\begin{equation}
\begin{split}
\Det(I-K_s+&\frac{1}{2} A_s\ket{\delta}\bra{1}A_s)\\
&=\exp\big(-\int_s^{+\infty} \rmd r \, (r-s)q_0(r)^2\big)\cosh\big(\frac{1}{2}\int_s^{+\infty} \rmd r \, q_0(r)\big)^2
\end{split}
\end{equation}
\item Orthogonal-thinned-like determinant
\begin{equation}
\label{eq:defOthinnedLike}
\begin{split}
\Det&(I-K_s-A_s\ket{\delta}\bra{1}(\sqrt{\alpha}\,  I-A_s))\\
&=\exp\big(-\int_s^{+\infty} \rmd r \, (r-s)q_0(r)^2\big)\bigg(\cosh\big(\int_s^{+\infty} \rmd r \, q_0(r)\big)-\sqrt{\alpha}\sinh\big(\int_s^{+\infty} \rmd r \, q_0(r)\big)\bigg)
\end{split}
\end{equation}
\end{itemize}
\end{result}

In the language of random matrix theory, orthogonal, unitary and symplectic determinants correspond to problems with Dyson index $\beta=1,2,4$. Recently, a random matrix problem related to the Tracy-Widom distribution with Dyson index $\beta=6$ was also brought in connexion with the functions $u_0$ and $q_0$ involved in this work in the case where the function $A$ is the standard Airy function, see Refs.~\cite{rumanov2016painleve, grava2016tracy}. It would be interesting to see if a rank-one or a rank-two perturbation of the Fredholm determinant of the Airy kernel could verify the same identities.
\subsection{Model-dependent equation }
All the above manipulations appear so far universal, hence the natural question  is to what extent the explicit expression of the function $A$ contributes to the problem. Recalling the hierarchy of equations
\begin{equation}
\begin{cases}
q_p'=q_{p+1}-q_0 u_p, \\
u_p'=-q_0 q_p,
\end{cases}
\end{equation}
we observe that the differential relations increase the index in the hierarchy by an increment of one for $q_p$, therefore at this stage, it is not possible to find a differential equation closing on the functions $\{ q_p, u_p \}$. Hence, we require at least another relation of the type
\begin{equation}
q_N=f(q_0,\dots, q_{N-1} ; u_0, \dots, u_{N-1})
\end{equation}
for some $N$ in order to have a differential equation of order $N$ on the function $q_0$ or $u_0$. We call this equation the \textit{model-dependent} equation and we will come back to it in Section~\ref{sec:MultiFermion} on the particular example of the Painlevé II hierarchy after having extended our current framework. \\

Traditionally, the literature has been mostly focused towards finding closed equations for the function $q_0$, as in the celebrated Painlevé II case related to the Airy kernel \cite{tracy1994level}. In the context of the Kardar-Parisi-Zhang equation, a number of work has drawn the attention to differential relations on $u_0$ and their connexion with the Kadomtsev–Petviashvili and Korteweg–de Vries equations, see Refs.~\cite{quastel2019kp, doussal2019large, claeys2020forth, barraquand2020half}.

\section{Inhomogeneous Fredholm determinants and general linear statistics}\label{sec:sigmaM}
The framework we have introduced so far can be extended to an inhomogeneous Fredholm determinant. To introduce the extension, let us briefly come back to the problem of linear statistics in random matrix theory or free fermionic theory.\\ 

An important property of determinantal point processes which we have not used so far is that any linear statistics of these points can be expressed as a Fredholm determinant. Indeed, for any function $\sigma$ supported on $\R$ and the determinantal point process $\lbrace a_i \rbrace$ defined uniquely by the correlation kernel $K_s$, we have 
\begin{equation}\label{eq:FredDet1}
\mathbb{E}\left[ \prod_{i=1}^\infty(1-\sigma(a_i))\right]=\Det\left( I-\sigma K_s \right)_{\mathbb{L}^2(\R)}
\end{equation}
The first part of this paper can be seen as a the particular case where $\sigma$ was a projector onto $\R_+$.

\begin{remark}
Averages such as the one in the left hand side of Eq.~\eqref{eq:FredDet1} often appear in the context of linear statistics. Quite generally, linear statistics problems consist in calculating the probability distribution of sums of the type $\mathcal{L}=\sum_i \phi(a_i)$, see Refs.~\cite{johansson2018gaussian, majumdar2011many, majumdar2012number, krajenbrink2018systematic, krajenbrink2019linear, grabsch2017truncated, krajenbrink2019beyond}. Evaluating the generating function of $\mathcal{L}$ amounts to averaging over products such as the one in the left hand side of Eq.~\eqref{eq:FredDet1}.
\end{remark}
\begin{remark}
In the language of free fermionic theories, the left hand side of Eq.~\eqref{eq:FredDet1} corresponds to an inhomogeneous full counting statistics or equivalently a quantum average of the observable $(1-\sigma(a_\ell))$ where $a_\ell$ is the position of the $\ell$-th fermion.
\end{remark}

In what follows, we will choose $\sigma$ to be increasing, smooth except at a finite number of points at which it has bounded jumps and with the following asymptotics\\
\begin{equation}
\lim_{t\to -\infty}\sigma(t)=0, \qquad \lim_{t\to +\infty} \sigma(t)=\gamma\in (0,1], \qquad \text{exponentially  fast}.
\end{equation}

Usually in the literature, the value $\gamma=1$ is only considered for the asymptotics, but here we also allow $0<\gamma\leq 1$ as in \cite{krajenbrink2019linear} so that our results also apply to study multiplicative statistics of thinned ensembles where eigenvalues are independently removed with probability $1-\gamma$. Decreasing $\gamma$ reduces the amount of repulsion between the eigenvalues, hence this parameter allows to study precise crossovers between correlated and uncorrelated statistics.\\

By Sylvester's identity, there are two equivalent representations of the Fredholm determinant \eqref{eq:FredDet1}. 
\begin{result}[Two representations for the Fredholm determinant]
We lift the operator $A_s$ to $ \mathbb{L}^2(\R) \to \mathbb{L}^2(\R_+)$ and denote its adjoint $A^\intercal_s$. Define the two operators $K_1:  \mathbb{L}^2(\R) \to \mathbb{L}^2(\R)$ and $K_2:  \mathbb{L}^2(\R_+) \to \mathbb{L}^2(\R_+)$ such that
\begin{equation}
\begin{split}
K_1&=\sigma A_s A_s^\intercal,\\
K_2&=A_s^\intercal\sigma A_s.
\end{split}
\end{equation}
Then we have
\begin{equation}
\Det(I-K_1)_{\mathbb{L}^2(\R)}=\Det(I-K_2)_{\mathbb{L}^2(\R_+)}.
\end{equation}
\end{result}

As before, since the operator $A_s$ has a kernel with an additive structure, there are a few calculation rules which we will use in the remaining. 
\begin{itemize}
\item For $K_1$
\begin{equation}
\label{eq:ruleK1}
\begin{split}
\partial_s K_1&=-\sigma A_s \ket{\delta}\bra{\delta} A_s^\intercal,\\
( \partial_x + \partial_y ) K_1 &=\partial_s K_1 + \sigma' A_sA_s^\intercal .
\end{split}
\end{equation}
\item For $K_2$
\begin{equation}
\label{eq:ruleK2}
\begin{split}
\partial_s K_2&=-A_s^\intercal \sigma' A_s,\\
( \partial_x + \partial_y ) K_2&=\partial_s K_2.
\end{split}
\end{equation}
\end{itemize}
Having defined the inhomogeneous extended framework, we introduce the new sequence of functions $\{ q_p,u_p \}$ as follows.
\begin{definition}[Generalization of the conjugated functions]
In the inhomogeneous setting, for any $s$ in $\R$ and $p$ in $ \N$,  we define the vector and matrix-valued functions
\begin{equation}
\label{eq:InhoQU}
q_p=A_s^{(p)}\frac{I}{I-K_2}\ket{\delta}, \qquad u_p=A_s^{(p)} \frac{I}{I-K_2}A_s^\intercal.
\end{equation}
where $A_s^{(p)}$ stands for the $p$-th derivative of $A_s$ with the convention $A_s^{(0)}=A_s$. The functions  $\{q_p\}$ are here interpreted as being column-like and $\{u_p \}$ matrix-like.  As noted by Bothner in \cite{bothner2020origins}, the map $s\mapsto q_p$ is smooth in light of its definition and the analyticity properties of the resolvent. Note that the definitions imply the following boundary conditions
\begin{equation}
q_p(t)\sim A^{(p)}(s+t), \qquad u_p(t,t')\sim \int_{\R_+}\rmd r \, A^{(p)}(t+r+s)A(r+t'+s)
\end{equation}
as $s\to +\infty$ for fixed $t,t'\in \R$.
\end{definition}
In terms of explicit integrals, Eqs.~\eqref{eq:InhoQU} are written as follows. Let $t,t' \in \R$,
\begin{equation}
\label{eq:InhoIntegralDef}
\begin{cases}
q_p(t)=\int_{\R_+} \rmd y \, A^{(p)}(t+s+y)(I-K_2)^{-1}(y,0)\\
u_p(t,t')=\iint_{\R_+^2}\rmd y \, \rmd z\, A(s+t+y)(I-K_2)^{-1}(y,z)A^{(p)}(z+t'+s)
\end{cases}
\end{equation}

To match the results of the Section~\ref{sec:lebesgueM}, we should choose $\sigma$ to be the projector on $\R_+$, so $\sigma'$ is the projector onto 0, $\sigma'=\ket{\delta}\bra{\delta}$. As we show next, this forces the variables $t,t'$ in Eqs.~\eqref{eq:InhoIntegralDef} to collapse to $0$ yielding back the definition of Eqs.~\eqref{eq:HomoDefInt}.\\

As in the homogeneous case, we present a $\tau$-representation of the Fredholm determinants. We will manipulate both operators $K_1$ and $K_2$ so to provide multiple equivalent representations and we will assume in the following that the resolvents of $K_1$ and $K_2$ exist, i.e. $I-K_1$ and $I-K_2$ are invertible. This can be proved similarly as Result~\ref{result:existenceResol} and this was done on a particular case in Ref.~\cite[Proposition~9.6]{bothner2020origins}. 
\begin{result}[$\tau$-representation of the inhomogeneous Fredholm determinants]
The following derivatives of the logarithm of the Fredholm determinants hold
\begin{itemize}
\item Using $K_1$, the first derivative reads
\begin{equation}
\label{eq:sig1logdet}
\partial_s \log \Det(I-K_1)= \bra{\delta} \frac{K_2}{I-K_2}\ket{\delta} .
\end{equation}
Introducing the canonical inner product on $\mathbb{L}^2(\R)$ for two functions $a,b$ and a measure $\sigma'$, $(a^\intercal\sigma' b)=\int_\R \rmd v \, a(v)\sigma'(v) b(v)$, the second derivative yields in terms of the function $q_0$
\begin{equation}
\label{eq:InhoD2}
\partial^2_s \log \Det(I-K_1)=-(q_0^\intercal \, \sigma' q_0),
\end{equation}
the notation $q_0^\intercal$ stands for the transpose of the vector-like function $q_0$.
\item Using $K_2$, the first derivative reads in terms of the function $u_0$
\begin{equation}
\label{eq:siglogdet}
\begin{split}
\partial_s \log \Det(I-K_2)= \Tr(\sigma' u_0).
\end{split}
\end{equation}
The second derivative is also given by Eq.~\eqref{eq:InhoD2}.
\end{itemize}
From the expression of the traces and the inner products, it is now clear that taking $\sigma'$ to be the projector to zero imposes $\{q_p, u_p\}$ to become scalar quantities 
\end{result}

\begin{proof}
\begin{enumerate}
\item Proof of \eqref{eq:sig1logdet}. Using the standard identity for the derivative of the logarithm of a determinant, the calculation rule related to $K_1$ \eqref{eq:ruleK1} and the cyclicity of the trace, we have
\begin{equation}
\begin{split}
\partial_s \log \Det(I-K_1)&= -\Tr(\frac{I}{I-K_1}\partial_s K_1)\\
&=\bra{\delta} A_s^\intercal \frac{I}{I-K_1}\sigma A_s \ket{\delta}\\
&= \bra{\delta} \frac{K_2}{I-K_2}\ket{\delta}.
\end{split}
\end{equation}
The third line connecting to the kernel $K_2$ is obtained by a series expansion term by term of the second line. 
\item Proof of \eqref{eq:InhoD2}.
From the formula $\partial_s (I-K_2)^{-1}=(I-K_2)^{-1} \partial_s K_2 (I-K_2)^{-1}$ and the derivative rule of $K_2$ \eqref{eq:ruleK2}, we obtain the second derivative as
\begin{equation}
\partial^2_s \log \Det(I-K_1)= -\bra{\delta} \frac{I}{I-K_2}A_s^\intercal \sigma' A_s \frac{I}{I-K_2}\ket{\delta} 
\end{equation}
Recognizing the appearance of  the function $q_0=A_s\frac{I}{I-K_2}\ket{\delta}$, we obtain that
\begin{equation}
\partial^2_s \log \Det(I-K_1)=-(q_0^\intercal \, \sigma' q_0).
\end{equation}
\item Proof of \eqref{eq:siglogdet}. From the standard identity for the derivative of the logarithm of a determinant and the calculation rule related to $K_2$ \eqref{eq:ruleK2} we obtain
\begin{equation}
\begin{split}
\partial_s \log \Det(I-K_2)&=\Tr(\frac{I}{I-K_2}A_s^\intercal \sigma' A_s)\\
&= \Tr(\sigma' u_0)
\end{split}
\end{equation}
where from the first to the second line we recognized the appearance of the matrix-like function $u_0$.
\item Proof of consistency \eqref{eq:siglogdet}\, =\, \eqref{eq:sig1logdet}. By Sylvester's identity, these two expressions should be equivalent, which we verify. From \eqref{eq:siglogdet}, we start by an integration by part on the product $A_s^\intercal \sigma' A_s$
\begin{equation}
\begin{split}
\Tr(\frac{I}{I-K_2}A_s^\intercal \sigma' A_s)&=-\Tr(\frac{I}{I-K_2}DA_s^\intercal \sigma A_s)-\Tr(\frac{I}{I-K_2}A_s^\intercal \sigma DA_s)\\
&= -\Tr(\frac{I}{I-K_2}DA_s^\intercal \sigma A_s)+\Tr(D\frac{I}{I-K_2}A_s^\intercal \sigma A_s)+\bra{\delta} \frac{K_2}{I-K_2}\ket{\delta} 
\end{split}
\end{equation}
From the first line to the second line, we proceeded to an integration by part on the product $DA_s (I-K_2)^{-1}$ and used the cyclicity of the trace. Finally, 
by cyclicity the first two terms cancel, yielding only the third one which is exactly \eqref{eq:sig1logdet}.
\end{enumerate}
\end{proof}

Bearing in mind the extended relation between the Fredholm determinant and the functions $q_0,u_0$, we present the integro-differential relations between the members of the inhomogeneous hierarchy.
\begin{result}[Integro-differential hierarchy of equations]
For all positive integer $p$, the following infinite recursion holds
\begin{equation}
\label{eq:IntegroHierarchyQU}
q_p'=q_{p+1}-u_p \sigma' q_0, \qquad  u_p'=-q_pq_0^\intercal.
\end{equation}
The prime ' stands for the derivative with respect to $s$.
\end{result}
One observation related to this integro-differential hierarchy is that the recursion on $q_p$ now becomes integral due to the presence of $\sigma'$ while the one on $u_p$ stays multiplicative. Furthermore, the derivative $u_p'$ is a rank-one operator for all $p$ as seen from the vector-like structure of $q_p$.\\

\begin{proof}
Starting from $q_p$, the first identity is obtained by differentiating the product of the two kernels and using the standard derivation formula for the resolvent.
\begin{equation}
\begin{split}
q_p'&=A_s^{(p+1)}\frac{I}{I-K_2}\ket{\delta}-A_s^{(p)}\frac{I}{I-K_2}A_s^\intercal \sigma' A_s \frac{I}{I-K_2}\ket{\delta}\\
&=q_{p+1}-u_p \sigma' q_0
\end{split}
\end{equation}
The second identity related to $u_p$ is more technical in this setting, since we do not have a generalisation of the Ferrari-Spohn derivative formula \eqref{lemma:FerrariSpohn}. We proceed by differentiating the product of the three kernels, use the calculation rule \eqref{eq:ruleK2}, proceed to various integration by parts on $\R_+$ with the operator notation $D^\intercal=-D-\ket{\delta}\bra{\delta}$ and use the commutation rule $[D,(I-K_2)^{-1}]=(I-K_2)^{-1}[D,K_2](I-K_2)^{-1}$.
\begin{equation}
\begin{split}
u_p'&=DA_s^{(p)} \frac{I}{I-K_2}A_s^\intercal+A_s^{(p)} \frac{I}{I-K_2}DA_s^\intercal+A_s^{(p)} \frac{I}{I-K_2}(DK_2+K_2D^\intercal) \frac{I}{I-K_2} A_s^\intercal\\
&= -A_s^{(p)} D\frac{I}{I-K_2}A_s^\intercal-A_s^{(p)}\ket{\delta}\bra{\delta} \frac{I}{I-K_2}A_s^\intercal+A_s^{(p)} \frac{I}{I-K_2}DA_s^\intercal+A_s^{(p)} \frac{I}{I-K_2}(DK_2+K_2D^\intercal) \frac{I}{I-K_2} A_s^\intercal\\
&= A_s^{(p)} \frac{I}{I-K_2} \left( K_2D-DK_2+DK_2+K_2D^\intercal \right)  \frac{I}{I-K_2} A_s^\intercal-A_s^{(p)}\ket{\delta}\bra{\delta} \frac{I}{I-K_2}A_s^\intercal\\
&= -A_s^{(p)} \frac{K_2}{I-K_2} \ket{\delta}\bra{\delta} \frac{I}{I-K_2} A_s^\intercal-A_s^{(p)}\ket{\delta}\bra{\delta} \frac{I}{I-K_2}A_s^\intercal\\
&= -A_s^{(p)} \frac{I}{I-K_2}\ket{\delta}\bra{\delta} \frac{I}{I-K_2} A_s^\intercal \\
&=-q_pq_0^\intercal
\end{split}
\end{equation}
\end{proof}
The hierarchy here again enjoys the existence of quadratic conserved quantities exhibiting here a much richer structure than the homogeneous counterpart since $\{q_p, u_p\}$ are not scalars but column and matrix-like.
\begin{result}[Symmetric and anti-symmetric conserved quantities]
For all $n$ in $\N$, the following quadratic quantities are invariant within the hierarchy
\begin{equation}
\label{eq:InhoIn}
\mathcal{I}_n=u_{2n+1}+u_{2n+1}^\intercal+\sum_{k=0}^{2n} (-1)^{k+1}[u_{2n-k}\sigma' u_k^\intercal- q_{2n-k}q_k^\intercal]=0
\end{equation}
and
\begin{equation}
\label{eq:InhoJn}
\mathcal{J}_n=u_{2n}-u_{2n}^\intercal+\sum_{k=0}^{2n-1} (-1)^{k+1}[u_{2n-k}\sigma' u_k^\intercal- q_{2n-k}q_k^\intercal]=0
\end{equation}
The sign $\!~^\intercal$ is meant as transpose. Note that the second equation is trivial in the inhomogeneous case $\sigma'=\ket{\delta }\bra{\delta}$ since all quantities become scalar and hence the left hand side \eqref{eq:InhoJn} is zero by parity of the summand. 
\end{result}
\begin{proof}
The proof is exactly the same as in the homogeneous case. Differentiating with respect to $s$ we obtain

\begin{equation}
\begin{split}
\mathcal{I}'_n&=-q_{2n+1}q_0^\intercal -q_{0}q_{2n+1}^\intercal -\sum_{k=0}^{2n} (-1)^{k+1}[q_{2n-k+1}q_k^\intercal+q_{2n-k}q_{k+1}^\intercal]\\
&=0
\end{split}
\end{equation}
and 
\begin{equation}
\begin{split}
\mathcal{J}'_n&=-q_{2n}q_0^\intercal +q_{0}q_{2n}^\intercal -\sum_{k=0}^{2n-1} (-1)^{k+1}[q_{2n-k}q_k^\intercal+q_{2n-1-k}q_{k+1}^\intercal]\\
&=0
\end{split}
\end{equation}
All quantities vanish as $s\to +\infty$, yielding no constant contribution upon integration.
\end{proof}

One could gather both conserved quantities onto a single equation introducing $(-1)^{n+1}$ term in front of $u_{n}^\intercal$ but we find it more instructive to exhibit the symmetric or anti-symmetric nature of the matrix like quantities.  For $n=0$, this result already appeared in the context of the finite-time solution to the Kardar-Parisi-Zhang equation with droplet initial condition in Ref.~\cite{ACQ} and more recently in \cite{bothner2020origins}. Its extension to arbitrary integer $n$ is new.\\

The discussion around the model-dependent equation stays valid in the inhomogeneous case since the ladder structure $p\to p+1$ is the same. To further understand how the constructed framework can be used practically, we will work in the following Section~\ref{sec:MultiFermion} on a specific example where we provide the model-dependent closure equation.

\section{Application to multi-critical fermions at finite temperature and the Painlevé II hierarchy}\label{sec:MultiFermion}
\subsection{The closure relation}
We now turn to an application of this work related to multi-critical fermions at finite-temperature in the physics community and the inhomogeneous Painlevé II hierarchy in the mathematics community. From now on, we will specify a function $A$ which we denote $\Ai_{2n+1}$, a higher-order Airy function, for $n$ in $\N$, satisfying the differential equation on $\R$
\begin{equation}
\label{eq:AiryDiff}
\Ai_{2n+1}^{(2n)}(x)=x \Ai_{2n+1}(x).
\end{equation}
In physics, these functions have appeared in multiple contexts, in random matrix models with external sources \cite{brezin1998level}, in lattice fermionic models long-range interactions and flat bands \cite{stephan2019free}, in multi-critical fermionic models \cite{le2018multicritical}.
We refer the reader to Ref.~\cite{le2018multicritical} for further physics insights on the problem.  In mathematics, these functions appear in the Painlevé II hierarchy, see Appendix~\ref{app:PIIhierarchy}, in the context of the steepest descent method where the saddle point enjoys a degeneracy and in the universality of limit shapes \cite{johansson2017edge}. Additional mathematical details on these functions, including their asymptotics, can be found in Refs.~\cite{le2018multicritical, cafasso2019fredholm}.\\

Coming back to our framework, the claim is that the differential relation \eqref{eq:AiryDiff} defines all the physics and the properties of the model and from it, we now show how to obtain the following closure, model-dependent, equation. We recall that we are interested in the Fredholm determinant $\Det(I-\sigma K_s)$ and the following quantities
\begin{equation}
\begin{split}
A_s(x,y)&=\Ai_{2n+1}(x+y+s)\\
K_2(x,y)&=\int_{\R} \rmd r  \, \sigma(r)\, \Ai_{2n+1}(x+s+r)\Ai_{2n+1}(y+s+r)\\
q_p&=A_s^{(p)}\frac{I}{I-K_2}\ket{\delta}\\
u_p&=A_s^{(p)} \frac{I}{I-K_2}A_s^\intercal
\end{split}
\end{equation}
The traditional choice of $\sigma$ to study multi-critical fermions at finite temperature \cite{le2018multicritical} is the so-called Fermi factor defined in terms of the inverse temperature $\beta$ as
\begin{equation}
\sigma(r)=\frac{1}{1+e^{-\beta r}}.
\end{equation}
At zero temperature, i.e. $\beta=+\infty$, the Fermi factor boils down to the indicator function on $\R_+$,  $\sigma(r)=\Theta(r)$. For the remainder of this Section, we will remain as general as possible with respect to the function $\sigma$ without imposing its explicit definition. 
In this setting, our first main result is the following. 
\begin{result}[Closure relation for multi-critical fermions at finite temperature]
For all integer $n$, the following identity for the function $q_{2n}$ related to the inhomogeneous Fredholm determinant defined from  $\Ai_{2n+1}$ holds
\begin{equation}
\label{eq:ClosureAiryHie}
q_{2n}=(s+X)q_0-\sum_{\ell=0}^{n-1}\left( u_{2n-1-2\ell}^\intercal \sigma' q_{2\ell}-u_{2n-2-2\ell}^\intercal \sigma' q_{2\ell+1} \right)
\end{equation}
where $X$ is the operator "multiplication by the left variable", i.e. for all $t \in \R$, $(Xq_0)(t)= t\, q_0(t)$.
\end{result}
For the choice $\sigma'=\ket{\delta}\bra{\delta}$, or equivalently the fermions at zero temperature, Eq.~\eqref{eq:ClosureAiryHie} yield back the results of \cite[Eq.~(110)]{doussal2018multicritical}.\\

\begin{proof}
Starting from the integral definition of $q_{2n}$ \eqref{eq:InhoIntegralDef}, for all $t$ in $\R$
\begin{equation}
\begin{split}
q_{2n}(t)&=\int_{\R_+} \rmd y \, \Ai_{2n+1}^{(2n)}(t+s+y)(I-K_2)^{-1}(y,0)\\
&=\int_{\R_+} \rmd y \, (t+s+y)\Ai_{2n+1}(t+s+y)(I-K_2)^{-1}(y,0)\\
\end{split}
\end{equation}
One separates the parenthesis in the integrand by the $y$-dependent and independent part and recognizes the operator $(s+X)q_0$ in the contribution $(t+s)\int_{\R_+} \rmd y \, \Ai_{2n+1}(t+s+y)(I-K_2)^{-1}(y,0)$. One needs to treat the remainder $\int_{\R_+} \rmd y \,y\, \Ai_{2n+1}(t+s+y) (I-K_2)^{-1}(y,0)$ and the standard trick is to interpret $y (I-K_2)^{-1}(y,0) $ as the commutator $[X,(I-K_2)^{-1}]$. Hence, so far we have obtained that
\begin{equation}
\begin{split}
q_{2n}&=(s+X)q_0 +A_s [X,\frac{I}{I-K_2}] \ket{\delta}\\
&=(s+X)q_0 +A_s \frac{I}{I-K_2}[X,K_2]\frac{I}{I-K_2} \ket{\delta}
\end{split}
\end{equation}
where we used the standard action of a commutator on the resolvent from the first to the second line.\\

Our task is now to understand how to treat the commutator $[X,K_2]$. A consequence of the differential equation for $\Ai_{2n+1}$ is the following identity for the kernel  $K_2$. For all $x,y$ in $\R_+$
\begin{equation}
(x-y)K_2(x,y)=(\partial_x^{2n}-\partial_y^{2n})K_2(x,y).
\end{equation}
Using the combinatorial equality for any integer $n$ and commuting quantities $a,b$,
\begin{equation}
\label{eq:CombiEq}
a^{2n}-b^{2n}=(a-b)\big(\sum_{\ell=0}^{n-1}a^{2n-2-2\ell}b^{2\ell}\big)(a+b),
\end{equation}
replacing $\partial_x$ by the operator $D$ and $\partial_y$ by $D^\intercal$ and finally recalling the rule for $K_2$ \eqref{eq:ruleK2}, $DK_2+K_2D^\intercal=-A_s^\intercal \sigma' A_s$, the commutator reads
\begin{equation}
[X,K_2]=-\sum_{\ell=0}^{n-1}\left((A_s^{(2n-1-2\ell)})^\intercal \sigma' A_s^{(2\ell)}-(A_s^{(2n-2-2\ell)})^\intercal \sigma' A_s^{(2\ell+1)}\right)
\end{equation}
Hence the remainder can now be expressed in terms of the functions $\{ q_p, u_p\}$ as 
\begin{equation}
\begin{split}
A_s\frac{I}{I-K_2}[X,K_2]\frac{I}{I-K_2}\ket{\delta}=-\sum_{\ell=0}^{n-1}\left( u_{2n-1-2\ell}^\intercal \sigma' q_{2\ell}-u_{2n-2-2\ell}^\intercal \sigma' q_{2\ell+1} \right)
\end{split}
\end{equation}
providing the rest of the model-dependent closure relation.
\end{proof}

\begin{remark}
One could wonder why we imposed an even number of derivatives in \eqref{eq:AiryDiff}. A crucial step in our calculation was the factorization of $a^{n}-b^{n}$ by $a+b$ which is possible only for even integers, hence the restriction. The case of odd derivatives is nonetheless also of interest since the Gaussian function $A'=-xA$ appears in the context of the real Ginibre ensemble of random matrices and the Zakharov-Shabat system, see Appendix~\ref{app:overview}, and the Pearcey function $A^{(3)}=xA$ appears in random matrix theory with external sources \cite{brezin2016random}. We leave these for a future work.
\end{remark}

In the remainder of this Section, we will investigate the precise integro-differential equations verified by $q_0$ for $n=1,2$ and discuss to some extent the outlook and perspectives of our results.

\subsection{The case $n=1$: the integro-differential Painlevé II equation}
For $n=1$, i.e. $A=\Ai$ the standard Airy function, using the closure relation \eqref{eq:ClosureAiryHie}, the first equations of the hierarchy  \eqref{eq:InhoQU} and the first conservation laws \eqref{eq:InhoIn}, \eqref{eq:InhoJn}, Amir, Corwin and Quastel obtained an integro-differential equation closed on $q_0$ for any function $\sigma$.

\begin{theorem}[First member of the integro-differential hierarchy {\cite[Proposition 5.2]{ACQ}}]
The function $q_0$ built from the Fredholm determinant with the Airy kernel and a measure $\sigma$ verifies the integro-differential Painlevé II equation
\begin{equation}
\label{eq:PIIACQ}
q_0''=(s+t)q_0+2q_0(q_0^\intercal \sigma' q_0)
\end{equation}
subject to the boundary condition  $q_0(t)\sim \Ai(s+t)$ as $s\to +\infty$ for fixed $t\in \R$. The prime $'$ stands for the derivative with respect to $s$.
\end{theorem}
When $\sigma$ is the projector onto $\R_+$ or equivalently when $\sigma'=\ket{\delta}\bra{\delta}$ is a Dirac mass onto 0, the integro-differential equation \eqref{eq:PIIACQ} boils down to the  Painlevé II equation
\begin{equation}
q_0''=sq_0+2q_0^3
\end{equation}
which is the standard result of Tracy and Widom \cite{tracy1994level}. The result of Amir, Corwin and Quastel was recently revisited by Bothner in \cite{bothner2020origins} by the means of Riemann-Hilbert methods. Note that it is still an open question to relate the integro-differential equation \eqref{eq:PIIACQ} to the results of Quastel and Remenik \cite{quastel2019kp} and Le Doussal \cite{doussal2019large}. We hope this article and the forthcoming work of Cafasso and Claeys \cite{claeys2020forth} will help bridge this gap.

\subsection{The case $n=2$: the integro-differential generalization of the second member of the Painlevé II hierarchy}

To the best of our knowledge, the case of general $n>1$ was only investigated in Refs.~\cite{le2018multicritical, cafasso2019fredholm} for the choice $\sigma$ to be taken equal to a projector onto $\R_+$ or equivalently for the fermions at zero temperature and the successive equations verified by $q_0$ were found to be the ones of the Painlevé II hierarchy, see Appendix~\ref{app:PIIhierarchy}. We extend this work to an arbitrary choice of $\sigma$ or equivalently to fermions at non-zero temperature and obtain for $n=2$ an integro-differential extension of the second member of the hierarchy as follows.

\begin{result}[Second member of the integro-differential hierarchy]
The function $q_0$ built from the Fredholm determinant with the kernel arising from the function $\Ai_5$ and a measure $\sigma$ verifies the integro-differential extension of the second equation of the Painlevé II hierarchy
\begin{equation}
\label{eq:PIIMoiMoi}
\begin{split}
q_0''''=&(s+t)  q_0+8  q_0' (q_0^\intercal \sigma'  q_0')+6
    q_0  (q_0^\intercal \sigma'  q_0'')-6
    q_0  (q_0^\intercal \sigma'  q_0)^2+2  q_0  ((q_0')^\intercal \sigma'  q_0')+4 q_0''  (q_0^\intercal \sigma'  q_0)
   \end{split}
\end{equation}
subject to the boundary condition  $q_0(t)\sim \Ai_5(s+t)$ as $s\to +\infty$ for fixed $t\in \R$. The prime $'$ stands for the derivative with respect to $s$.

\end{result}

The derivation of this equation is mostly technical and cumbersome. It solely involves the hierarchy of equations \eqref{eq:IntegroHierarchyQU}, the conservation laws \eqref{eq:InhoIn}, \eqref{eq:InhoJn} and the closure relation \eqref{eq:ClosureAiryHie}. We describe the ad-hoc procedure to obtain \eqref{eq:PIIMoiMoi} as follows\\

\begin{proof}[Procedure]
\begin{enumerate}
\item Differentiate $q_0$ four times;
\item Replace the value of $q_4$ by the closure relation;
\item Use the flow invariance for $u_3+u_3^\intercal$;
\item Use the flow invariance for $u_2-u_2^\intercal$;
\item Replace $q_2$ by $q_1'+u_1\sigma'q_0$;
\item Use the symmetry of the inner-product $(v^\intercal u_1 v)=\frac{1}{2}(v^\intercal [u_1+u_1^\intercal] v)$ and the flow invariance for $u_1+u_1^\intercal$;
\item Replace $q_1$ by $q_0'+u_0\sigma'q_0$; 
\item Use that $u_0=u_0^\intercal$;
\item Use the symmetry of the various inner-products $(v^\intercal w)=(w^\intercal v)$.
\end{enumerate}
\end{proof}
When $\sigma$ is the projector onto $\R_+$ or equivalently when $\sigma'=\ket{\delta}\bra{\delta}$ is a Dirac mass onto 0 or when we consider the fermions at zero temperature, the integro-differential equation \eqref{eq:PIIMoiMoi} reduces to the second equation of the Painlevé II hierarchy, see \cite{doussal2018multicritical} and Appendix~\ref{app:PIIhierarchy}.
\begin{equation}
\label{eq:PIIsecondMember}
q_0''''= sq_0+10q_0 (q_0')^2+10q_0^2q_0''-6q_0^5 
\end{equation}
By construction, we have shown the existence of the solution the integro-differential equation \eqref{eq:PIIMoiMoi} with the prescribed boundary condition.  \\

We now provide a few interpretation, comments and outlooks on our construction and results.
\begin{enumerate}
\item Assuming the whole Painlevé II hierarchy can be continued to its integro-differential counterpart through our framework (with a closed integro-differential equation on $q_0$), we notice an important structure in the equations. Since $q_0$ is vector-like, to obtain equations such as \eqref{eq:PIIACQ} and \eqref{eq:PIIMoiMoi} with a vector-like structure, only odd powers of $q_0$ can arise. One power of $q_0$ to retain the vector-like structure and even powers of $q_0$ involved in the inner products $(q_0^\intercal \sigma' q_0)$ with arbitrary derivatives taken on the $q_0$'s. This remark extends to the usual Painlevé II hierarchy (seen as a special case $\sigma'\to \ket{\delta}\bra{\delta}$) and provides an interpretation to why only odd powers of $q_0$ appear.
\item Furthermore, a few other cyclic permutations of the $q_0$ and its derivatives in \eqref{eq:PIIMoiMoi} could have led to \eqref{eq:PIIsecondMember} upon the choice $\sigma'\to \ket{\delta}\bra{\delta}$. It would interesting to know if the precise coefficients in $\eqref{eq:PIIMoiMoi}$ have a combinatorial interpretation and whether they are related to integrability or solvability. 
\item It would be of reasonable importance to obtain a more systematic way to derive in our framework the integro-differential equations of the Painlevé II hierarchy than the procedure \textit{"by hand"}. This point was already raised in \cite{le2018multicritical} and we leave this open. 
\item At first glance, it might look surprising to relate integro-differential equations to Fredholm determinants with an \textit{inhomogeneous measure} $\sigma$ (also interpreted as an inhomogeneous counting statistics in a fermionic theory). In light of integrable models in physics, this is not unfamiliar since for instance the homogeneous Heisenberg spin model maps to the attractive nonlinear Schrödinger equation and the inhomogeneous Heisenberg model maps to an integro-differential version of the nonlinear Schrödinger equation, see Ref.~\cite{balakrishnan1982inhomogeneous}. 
\item Since some of the problems considered within this work are also amenable by Riemann-Hilbert methods, see \cite{cafasso2019fredholm,bothner2020origins}, it would interesting to see how integro-differential manipulations would emerge in that framework more systematically.
\item There is an additional duality in the context of multi-critical fermions that we have not explored in this Section. We have seen that investigating the Fredholm determinants $\Det(I-\sigma A_s A_s)$ and $\Det(I- A_s \sigma A_s)$ is strictly equivalent on a mathematical basis. The first determinant has the physical interpretation of a linear statistics for a determinantal process with the kernel $A_s^2$ while the second has the interpretation of a counting statistics for a determinantal process with kernel $A_s \sigma A_s$. It is then natural to extend the linear statistics problem to the determinantal process with kernel $A_s \sigma A_s$, hence studying a Fredholm determinant of the type $\Det(I-\mu A_s \sigma A_s)$ for two different measures $\mu, \sigma$. We expect this type of kernel to also generate a hierarchy of functions $\{ q_p, u_p\}$ and  we leave this question for a future work.
\end{enumerate}

\begin{remark} Let us provide a final remark on the Painlevé II hierarchy. In the same spirit as Ref.~\cite{cafasso2019fredholm}, we can obtain a slight generalization of the above closure relation \eqref{eq:ClosureAiryHie}. Let $A$ verify the following differential relation for $n \in \N$, $\tau_1, \dots,\tau_{n-1} \in \R$ and $x \in \R$,
\begin{equation}
\partial_x^{2n} A(x)+\sum_{k=1}^{n-1} \tau_k \partial_x^{2k} A(x)=xA(x).
\end{equation}
Defining $\tau_n=1$, the combinatorial identity replacing \eqref{eq:CombiEq} is in this case
\begin{equation}
\sum_{k=1}^n \tau_k (a^{2k} -b^{2k})=(a-b)\big[\sum_{k=1}^n \tau_k \sum_{\ell=0}^{k-1} a^{2k-2-2\ell}b^{2\ell}\big](a+b)
\end{equation}
and the closure relation replacing \eqref{eq:ClosureAiryHie} reads
\begin{equation}
q_{2n}+\sum_{k=1}^{n-1} \tau_k q_{2k}=Xq_0-\sum_{k=1}^n \tau_k \sum_{\ell=0}^{k-1}\left( u_{2k-1-2\ell}^\intercal \sigma' q_{2\ell}-u_{2k-2-2\ell}^\intercal \sigma' q_{2\ell+1} \right).
\end{equation}
We expect that a closed differential equation on $q_0$ will also exist and all above conclusions will remain valid.
\end{remark}

\section{Application to the exact solution of the Zakharov-Shabat system} \label{sec:ZS}
In this Section, we propose to study the Zakharov-Shabat  system \cite{ablowitz1991solitons, shabat1972exact} and unveil a connexion with the family of Fredholm determinants considered in this work. To that aim, we first introduce the results of \cite{baik2020largest} and we will expand them to reveal a one to one correspondence between the Fredholm determinant \eqref{eq:FredDet} and the exact solution to the Zakharov-Shabat system. 

\subsection{The Zakharov-Shabat system and the largest real eigenvalue of the real Ginibre ensemble}
The beginning of this Section follows closely the presentation and the results of Ref.~\cite{baik2020largest} upon the choice of conventions closer to the one used for the Gaussian ensembles in Appendix~\ref{subsec:appGaussian} and the elliptic ensembles in Appendix~\ref{subsec:appEllipticGinibre}. Consider a matrix $M$ of size $N\times N$ belonging to the real Ginibre ensemble, that is the entries of $M$ are independent and identically distributed according to a Gaussian probability distribution
$\tilde{f}(M_{ij})\propto e^{-\frac{N}{2}M_{ij}^2}$, so that the distribution of the matrix $M$ is given by 
\begin{equation}
f(M)=\prod_{i,j=1}^N\tilde{f}(M_{ij})\propto e^{-\frac{N}{2}\Tr(MM^\intercal)}.
\end{equation}
Then, in our system of notations, we have the following result from Rider, Sinclair and from Poplavskyi, Tribe and Zaboronski relating the cumulative distribution of the largest real eigenvalue of the real Ginibre ensemble to a Fredholm determinant with a scalar-valued kernel. 
\begin{theorem}[Rider, Sinclair \cite{rider2014extremal}, 2014; Poplavskyi, Tribe, Zaboronski \cite{poplavskyi2017distribution}, 2017] Denoting $\{ z_j(M)\}_{j=1}^N$ the complex eigenvalues of $M$, we have for $s$ in $\R$
\begin{equation}\label{e:1}
	\lim_{N\rightarrow\infty}\mathbb{P}\left(\frac{\max_{j:z_j\in\mathbb{R}}z_j(M)-1}{N^{-1/2}}\leq s\right)=\sqrt{\det\big(I-K_s-A_s\ket{\delta}\bra{1}(I-A_s)\big)_{\mathbb{L}^2(\R_+)}},
\end{equation}
where $K_s:\mathbb{L}^2(\R_+)\rightarrow \mathbb{L}^2(\R_+)$ is the operator with kernel
\begin{equation}\label{e:2}
K_s(x,y)=\frac{1}{\pi}\int_0^{\infty}e^{-(x+u+s)^2}e^{-(y+u+s)^2}\,\rmd u.
\end{equation}
and, defining the Gaussian $A(x)=\frac{1}{\sqrt{\pi}}e^{-x^2}$, $A_s: \mathbb{L}^2(\R_+)\to \mathbb{L}^2(\R_+)$ is the operator with kernel $A_s(x,y)=A(s+x+y)$. In particular, we have that $K_s=A_s^2$.
\end{theorem} 

So to include this problem into our framework, let us mention that although the operator $A_s$ does not have the orthogonality relation
\begin{equation}
\int_\R \rmd r A_s(x,r)A_s(r,y)=\frac{1}{\sqrt{2\pi}}e^{-\frac{(x-y)^2}{2}}\neq \delta(x-y)
\end{equation}
it was shown in Ref.~\cite[Lemma 2.1]{baik2020largest} that $K_s$ is bounded from above by the identity and that $I-\gamma K_s$ is invertible on $\mathbb{L}^2(\R_+)$ for all $\gamma \in [0,1]$. This ensures that the construction of the functions $\{ q_p, u_p \}_{p\in \N}$ is well-posed and thus all the results are valid. Hence our framework allows to define for $\gamma \in [0,1]$ 
\begin{equation}
q_0(s)=\sqrt{\gamma }\bra{\delta}\frac{A_s}{I-\gamma K_s}\ket{\delta}, \quad \text{so that} \quad q_0(s)^2=-\frac{\rmd^2}{\rmd s^2}\log \Det[I-\gamma K_s].
\end{equation}

Baik and Bothner have extended the Fredholm determinant \eqref{e:1} to an orthogonal thinned-like version
\begin{equation}
\label{eq:GinibreOEthinned}
F(s,\gamma)=\sqrt{\det\big(I-\gamma K_s-\gamma A_s\ket{\delta}\bra{1}(I-A_s)\big)_{\mathbb{L}^2(\R_+)}}
\end{equation}
and have proved the following relation between $F(s,\gamma)$ and a distinguished solution to the Zakharov-Shabat system
\begin{theorem}[Baik, Bothner {\cite{baik2020largest}}, 2020]\label{main1}
For any $(s,\gamma)\in\mathbb{R}\times[0,1]$,
\begin{equation}\label{e:6}
	\big(F(s;\gamma)\big)^2=\exp\bigg[-\frac{1}{4}\int_s^{\infty}(x-s)\left|y\left(\frac{x}{2};\gamma\right)\right|^2\rmd x\bigg]\bigg\{\cosh\mu(s;\gamma)-\sqrt{\gamma}\sinh\mu(s;\gamma)\bigg\},
\end{equation}
using the abbreviation
\begin{equation*}
	\mu(s;\gamma):=-\frac{\I}{2}\int_s^{\infty}y\left(\frac{x}{2};\gamma\right)\,\rmd x,
\end{equation*}
and where $y=y(x;\gamma):\mathbb{R}\times[0,1]\rightarrow\I\mathbb{R}$ equals $y(x;\gamma):=2\I X_1^{12}(x,\gamma)$ in terms of the matrix coefficient ${\bf X}_1(x,\gamma)$ in condition (3) of Riemann-Hilbert problem (RHP) \ref{master0} below.
\end{theorem}
Before reviewing the RHP considered in Theorem~\ref{main1}, let us state that their result matches ours as follows: for all real $s$ we have $q_0(s)=\frac{1}{2\I} y\left(\frac{s}{2}; \gamma\right) $ and the determinant \eqref{eq:GinibreOEthinned} is orthogonal-thinned-like as defined in \eqref{eq:defOthinnedLike}. This confirms easily that the following asymptotics holds $ y(s;\gamma)=_{s\to +\infty}2\I \sqrt{ \frac{ \gamma }{\pi}}e^{-4s^2}$. Now that the connexion between the two results is established, let us recall the definition of the corresponding RHP and then let mention the mapping to the Zakharov-Shabat system.

\begin{definition}[Riemann-Hilbert problem]\label{master0} For any $(x,\gamma)\in\mathbb{R}\times[0,1]$, determine ${\bf X}(z)={\bf X}(z;x,\gamma)\in\mathbb{C}^{2\times 2}$ such that
\begin{enumerate}
	\item ${\bf X}(z)$ is analytic for $z\in\mathbb{C}\setminus\mathbb{R}$ and has a continuous extension on the closed upper and lower half-planes.
	\item The limiting values ${\bf X}_{\pm}(z)=\lim_{\epsilon\downarrow 0}{\bf X}(z\pm\I\epsilon),z\in\mathbb{R}$ satisfy the jump condition
	\begin{equation}\label{djump}
		{\bf X}_+(z)={\bf X}_-(z)\begin{bmatrix}1-|r(z)|^2 & -\bar{r}(z)e^{-2\I xz}\smallskip\\ r(z)e^{2\I xz} & 1\end{bmatrix},\ \ z\in\mathbb{R}\ \ \ \ \ \textnormal{with}\ \ \ r(z)=r(z;\gamma)=-\I\sqrt{\gamma}\,e^{-\frac{1}{4}z^2}.
	\end{equation}
	\item As $z\rightarrow\infty$, we require the normalization
	\begin{equation}\label{eq:RiemannLaurentExp}
		{\bf X}(z)=\mathbb{I}+{\bf X}_1z^{-1}+{\bf X}_2z^{-2}+\mathcal{O}\big(z^{-3}\big);\ \ \ \ \ {\bf X}_i={\bf X}_i(x,\gamma)=\big[X_i^{jk}(x,\gamma)\big]_{j,k=1}^2.
	\end{equation}
\end{enumerate}
\end{definition}
It was then shown in \cite{baik2020largest} that this RHP is then translated easily in the language of the Zakharov-Shabat system as follows. 
\begin{corollary}[Relation to the Zakharov-Shabat system. Baik, Bothner {\cite{baik2020largest}}, 2020]
The quantity
\begin{equation}\label{eq:PsiXrelation}
	{\bf\Psi}(z):={\bf X}(z)e^{-\I xz\sigma_3},\ \ \ z\in\mathbb{C}\setminus\mathbb{R}; \qquad \sigma_3=\begin{bmatrix}1 & 0\\ 0 & -1\end{bmatrix}
\end{equation}
solves a RHP with an $x$-independent jump on $\mathbb{R}$, thus $\frac{\partial{\bf\Psi}}{\partial x}{\bf \Psi}^{-1}$ is an entire function. In fact, using condition (3) in RHP \ref{master0} and Liouville's theorem, ${\bf \Psi}$ verifies 
\begin{equation}\label{ZS:1}
	\frac{\partial{\bf \Psi}}{\partial x}=\left\{-\I z\sigma_3+\begin{bmatrix}0 & y\\ \bar{y} & 0\end{bmatrix}\right\}{\bf\Psi}
\end{equation}
This is the celebrated first order system, known as Zakharov-System system \cite{ablowitz1991solitons, shabat1972exact}. Hence, the quantity \eqref{e:6} depends on a distinguished solution $y(x;\gamma)$ of the inverse scattering transform for the Zakharov-Shabat system \eqref{ZS:1} subject to the reflection coefficient $r(z;\gamma)=-\I\sqrt{\gamma}\,e^{-\frac{1}{4}z^2}$.
\end{corollary}
From the work of Baik and Bothner, Ref.~\cite[Eq. (3.25)]{baik2020largest}, we further have the direct identification
\begin{equation}
\label{eq:app:indentif}
\begin{cases}
q_0=X_1^{12}(\frac{s}{2}, \gamma) \\
u_0=\I X_1^{11}(\frac{s}{2}, \gamma)\\
\end{cases} \quad 
\Longrightarrow \qquad 
{\bf X}_1(\frac{s}{2},\gamma)=
\begin{bmatrix}
-\I u_0 &q_0\smallskip\\ 
q_0 & \I u_0\end{bmatrix}
\end{equation}
In order to go beyond the first order of the Laurent expansion \eqref{eq:RiemannLaurentExp}, we study the differential relation verified by the solution to the Riemann-Hilbert problem. From the definition \eqref{eq:PsiXrelation} and the Zakharov-Shabat system \eqref{ZS:1}, the matrix ${\bf X}$ obeys the differential equation 
\begin{equation}
\label{eq:ZakharovShabatX}
\frac{\partial {\bf X}}{\partial x}=\I z [{\bf X}, \sigma_3 ]+2\I  \begin{bmatrix}
0 &q_0 \smallskip\\ 
-q_0 & 0 \end{bmatrix}{\bf X}
\end{equation}
where $[ \cdot, \cdot]$ is the standard commutator. A direct induction shows that our hierarchy $\{ q_p, u_p \}$ \eqref{eq:ConjVar}
is directly related to the general coefficients of the Laurent expansion \eqref{eq:RiemannLaurentExp} where we denote the coefficient of order $z^{-p}$ by ${\bf X}_{p}$. 
\begin{result}[Explicit Laurent series for the RHP \ref{master0}]
For all $p$ in $\N$, the expression of the $p+1$-th coefficient of the Laurent series is 
\begin{equation}
{\bf X}_{p+1}(\frac{s}{2},\gamma)=
\begin{bmatrix}
(-\I)^{p+1} u_p &\I^{p} q_p\smallskip\\ 
(-\I)^{p} q_p & \I^{p+1} u_p\end{bmatrix}
\end{equation}
\end{result}
\begin{proof}
The proof results from the expansion of \eqref{eq:ZakharovShabatX} in powers of $1/z$ around infinity so that the recursion is given by 
\begin{equation}
\frac{\partial {\bf X}_p}{\partial x }=\I  [{\bf X}_{p+1}, \sigma_3 ]+2\I  \begin{bmatrix}
0 &q_0 \smallskip\\ 
-q_0 & 0 \end{bmatrix}{\bf X}_p
\end{equation}
Initiating the recursion from the identification \eqref{eq:app:indentif} and using the hierarchy of equation \eqref{eq:HomoQuasiUniversalHie} 
\begin{equation}
\begin{cases}
q_p'=q_{p+1}-q_0 u_p, \\
u_p'=-q_0 q_p,
\end{cases}
\end{equation}
with the change $x\to s/2$ leads directly to the result. 
\end{proof}
The Laurent series is therefore given formally as
\begin{equation}
{\bf X}(z)=\mathbb{I}+\sum_{\ell=1}^\infty \frac{1}{z^\ell }\begin{bmatrix}
(-\I)^{\ell} u_{\ell-1} &\I^{\ell-1} q_{\ell-1}\smallskip\\ 
(-\I)^{\ell-1} q_{\ell-1} & \I^{\ell} u_{\ell-1}\end{bmatrix}
\end{equation}
Recalling the definitions of $\{ q_p, u_p\}$ from Eq.~\eqref{eq:ConjVar}
\begin{equation}
q_p=\bra{\delta} \frac{A_s^{(p)}}{I-K_s}\ket{\delta}, \qquad u_p=\bra{\delta} A_s\frac{I}{I-K_s}A_s^{(p)}\ket{\delta},
\end{equation}
we recast formally ${\bf X}(z)$ as 
\begin{equation}
\label{eq:LaurentSeriesXRiemann}
{\bf X}(z)=\mathbb{I}+\bra{\delta}\sum_{\ell=1}^\infty \frac{D^{\ell-1}}{z^\ell }\begin{bmatrix}
(-\I)^{\ell}A_s &\I^{\ell-1} \smallskip\\ 
(-\I)^{\ell-1}  & \I^{\ell} A_s\end{bmatrix}
 \frac{A_s}{I-K_s}\ket{\delta}
\end{equation}
where we recall that $D$ is the derivative operator. It is tempting to sum the geometric series in \eqref{eq:LaurentSeriesXRiemann}. We now formally do so and show that it provides the explicit unique solution to the Riemann-Hilbert problem~\ref{master0}.
\begin{result}[Explicit solution to the Riemann-Hilbert problem in the complex plane]\label{result:explicitNewSolution}
The Hilbert Boundary Value Problem \ref{master0} admits the following solution for $z\in \C$ with the identification $x=s/2$
\begin{equation}\label{eq:explicitNewSolutionRiemann}
{\bf X}(z)=\mathbb{I}+\bra{\delta}\begin{bmatrix}
-\frac{1}{D-\I z}A_s  &\frac{\I}{D+\I z} \smallskip\\ 
-\frac{\I}{D-\I z}  &- \frac{1}{D+\I z} A_s  \end{bmatrix}
 \frac{A_s}{I-K_s}\ket{\delta}
\end{equation}
where $D$ is the derivative operator and the formal notation $1/(D\pm \I z)$ is meant as the resolvent of the derivative operator.
\end{result}
Before showing that our solution verifies the jump condition \eqref{djump} and has the right analytic properties, we shall look at an interesting feature of ${\bf X}$ which is its value at $z=0$. Using that integral relation $\bra{\delta}D^{-1}=-\bra{1}$ and the $\tau$-representation of the function $b$ defined in \eqref{eq:FerSp} and \eqref{eq:FerrSpohnPainlevQ}, we obtain
\begin{equation}
\begin{split}
{\bf X}(0)&=\mathbb{I}+\bra{1}\begin{bmatrix}
A_s &-\I  \smallskip\\ 
\I    & A_s \end{bmatrix}
\frac{A_s}{I-K_s}\ket{\delta}\\
 &= \bra{1}\begin{bmatrix}
1 &-\I A_s \smallskip\\ 
\I A_s    & 1 \end{bmatrix}
\frac{I}{I-K_s}\ket{\delta}\\
&=  \begin{bmatrix}
\cosh(-\int_s^\infty q_0) &\I \sinh(-\int_s^\infty q_0) \smallskip\\ 
-\I \sinh(-\int_s^\infty q_0)    & \cosh(-\int_s^\infty q_0) \end{bmatrix}
 \end{split}
\end{equation}
Note that $X(0)$ verifies the first order Taylor expansion of \eqref{eq:ZakharovShabatX}, i.e.  $
\partial_s X(0) =\I \begin{bmatrix}
0 &q_0 \smallskip\\ 
-q_0 & 0 \end{bmatrix}  X(0)$.\\

The resolvent of the derivative operator can be explicitly obtained and hence this allows us to simplify the expression of ${\bf X}$. Indeed, for any sufficiently smooth functions $f,g$, we have for $z\in \C$
\begin{itemize}
\item for $\Im(z)>0$,
\begin{equation}\label{eq:resolvInverse1}
\partial_x f+ \I z f=g \quad \Leftrightarrow \quad f(x)=-\int_0^{+\infty}\rmd t \,  g(x+t) e^{ \I z t}
\end{equation}
\item for $\Im(z)<0$,
\begin{equation}\label{eq:resolvInverse2}
\partial_x f+ \I z f=g \quad  \Leftrightarrow \quad f(x)=\int_\R g(x+t)e^{\I zt} -\int_0^{+\infty}\rmd t \,  g(x+t) e^{ \I z t}
\end{equation}
\end{itemize}

Since we constrain the boundary by $\bra{\delta}$ in Eq.~\eqref{eq:explicitNewSolutionRiemann}, we denote the resolvent of the derivative operator as follows:
\begin{itemize}
\item for $\Im(z)>0$ from \eqref{eq:resolvInverse1}
\begin{equation}
\bra{\delta} \frac{I}{D+ \I z}A_s =-\bra{1} e^{ \I z X}  A_s
\end{equation}
where $X$ is the operator \textit{"multiplication by the left variable"}.
\item for $\Im(z)<0$ from \eqref{eq:resolvInverse2}
\begin{equation}
\bra{\delta} \frac{I}{D+ \I z}A_s =\big( \int_\R \rmd u \, e^{ \I z u}A(u) \big) \bra{1}e^{- \I z (s+X)}-\bra{1} e^{ \I z X}  A_s
\end{equation}
The reason for the first term of the right-hand side is that the kernel of $A_s$ has an additive structure $A_s(x,y)=A(x+y+s)$ and hence $\int_\R \rmd u A_s(u,y)e^{\I z u}=(\int_\R \rmd u \, e^{ \I z u}A(u)  )e^{-\I (s+y)}$.\\
\end{itemize}

Equipped with the evaluation of the resolvent of the derivative operator and using the value of the inner product $\bra{1}e^{\pm \I z X}\ket{\delta}=1$, we deduce the following representation for ${\bf X}$.
\begin{itemize}
\item For $\Im(z)>0$ 
\begin{equation}
\label{eq:expressionXupperplane}
\begin{split}
{\bf X}(z) =\bra{1}&\begin{bmatrix}
e^{-\I z X}  &-\I e^{\I z X} A_s\smallskip\\ 
\I e^{-\I z X}A_s  &e^{\I z X}  \end{bmatrix}
 \frac{I}{I-K_s}\ket{\delta}\\
 &-\big(\int_\R \rmd u \, e^{- \I z u}A(u)\big) \bra{1}e^{ \I z (s+X)}
\begin{bmatrix}
 A_s &0\smallskip\\ 
\I  &0 \end{bmatrix} \frac{I}{I-K_s}\ket{\delta}\\
 \end{split}
\end{equation}
\item For $\Im(z)<0$, 
\begin{equation}\label{eq:expressionXlowerplane}
\begin{split}
{\bf X}(z) =\bra{1}&\begin{bmatrix}
e^{-\I z X}  &-\I e^{\I z X} A_s\smallskip\\ 
\I e^{-\I z X}A_s  &e^{\I z X}  \end{bmatrix}
 \frac{I}{I-K_s}\ket{\delta}\\
 &-\big( \int_\R \rmd u \, e^{ \I z u}A(u) \big) \bra{1}e^{- \I z (s+X)}
\begin{bmatrix}
0&-\I   \smallskip\\ 
0&    A_s \end{bmatrix} \frac{I }{I-K_s}\ket{\delta}\\
 \end{split}
\end{equation}
\end{itemize}

Since $A_s$ is Hilbert-Schmidt and using the analyticity properties of the resolvent of $K_s$, it is clear from \eqref{eq:expressionXupperplane} and \eqref{eq:expressionXlowerplane} that ${\bf X}(z)$ is analytic for $z\in\mathbb{C}\setminus\mathbb{R}$ and has a continuous extension on the closed upper and lower half-planes. Hence the first condition of the Riemann-Hilbert problem \ref{master0} is verified. We will now show that the jump condition is as well obeyed by our solution. Quite surprisingly, the reflection coefficient is easily related to the function $A$ as follows.

\begin{definition}[Reflection coefficient]
We define the reflection coefficient as 
\begin{equation}
r(z)=-\I \int_\R \rmd u \, e^{- \I z u}A(u) 
\end{equation}
If $A(u)=\sqrt{\frac{\gamma}{\pi}}e^{-u^2}$ then $r(z)=-\I \sqrt{\gamma}e^{-\frac{1}{4}z^2}$, matching the definition of the reflection coefficient of the Zakharov-Shabat system related to the largest real eigenvalue of the real Ginibre ensemble \cite{baik2020largest}.
\end{definition}

\begin{result}[Jump condition]
The matrix ${\bf X}$ of Result~\ref{result:explicitNewSolution} verifies the jump condition \eqref{djump}.
\end{result}
\begin{proof}
The final step is the evaluation of the values ${\bf X}_{\pm}(z)=\lim_{\epsilon\downarrow 0}{\bf X}(z\pm\I\epsilon),z\in\mathbb{R}$ from Eqs.~\eqref{eq:expressionXlowerplane}, \eqref{eq:expressionXupperplane} and the definition of the reflection coefficient.
\begin{itemize}
\item For $\Im(z)>0$
\begin{equation}
\begin{split}
{\bf X}_+(z)
 &=\bra{1}\left\{ \begin{bmatrix}
e^{-\I z X}  &-\I e^{\I z X} A_s\smallskip\\ 
\I e^{-\I z X}A_s  &e^{\I z X}  \end{bmatrix}
-\I r(z)e^{\I z (X+s)}
\begin{bmatrix}
 A_s &0\smallskip\\ 
\I &0 \end{bmatrix}\right\} \frac{I}{I-K_s}\ket{\delta}\\
 \end{split}
\end{equation}
\item for $\Im(z)<0$
\begin{equation}
\begin{split}
{\bf X}_-(z) &=\bra{1}\left\{ \begin{bmatrix}
e^{-\I z X}  &-\I e^{\I z X} A_s\smallskip\\ 
\I e^{-\I z X}A_s  &e^{\I z X}  \end{bmatrix}
+\I \bar{r}(z)e^{-\I z (X+s)}
\begin{bmatrix}
0&-\I  \smallskip\\ 
0&   A_s \end{bmatrix} \right\}\frac{I}{I-K_s}\ket{\delta}\\
 \end{split}
\end{equation}
\end{itemize}
The matrices inside the braces verify the required jump identity 
\begin{equation}
\begin{split}
&\begin{bmatrix}
e^{-\I z X} -\I r(z)e^{\I z(X+s)}A_s &-\I e^{\I z X} A_s\smallskip\\ 
\I e^{-\I z X}A_s +r(z)e^{\I z(X+s)} &e^{\I z X}  \end{bmatrix}
=\\
& \begin{bmatrix}
e^{-\I z X}  &-\I e^{\I z X} A_s+ \bar{r}(z) e^{-\I z(X+s)}\smallskip\\ 
\I e^{-\I z X}A_s  &e^{\I z X} +\I \bar{r}(z)e^{-\I z(X+s)}A_s \end{bmatrix}
\begin{bmatrix}1-|r(z)|^2 & -\bar{r}(z)e^{-\I sz}\smallskip\\ r(z)e^{\I sz} & 1\end{bmatrix}
\end{split}
\end{equation}
 \end{proof}
Since the Riemann-Hilbert problem \ref{master0} is uniquely solvable, see Ref.~\cite{bothner2020origins}, we have therefore found its explicit solution. Quite interestingly, none of our manipulations depended on the precise expression of the function $A$, hence we have obtained a more general solution to the inverse scattering transform for the Zakharov-Shabat system. We have unveiled the precise relation between the quasi-universal hierarchy \eqref{eq:HomoQuasiUniversalHie} and the Riemann-Hilbert problem.

\subsection{Explicit solution to the inverse scattering transform for the Zakharov-Shabat system}
Our above manipulations yield a new explicit solution to the Zakharov-Shabat system for a family of reflection coefficients. Let $A$ a real-valued function, $\gamma$ in $[0,1]$ and $s$ in $\R$ such that the integral operator $A_s: \mathbb{L}^2(\R_+)\to \mathbb{L}^2(\R_+)$ with kernel $A_s(x,y)=\sqrt{\gamma} A(s+x+y)$ is Hilbert-Schmidt and the operator $K_s=A_s^2$ is bounded by above by the identity. Consider the Riemann-Hilbert problem as follows

\begin{definition}\label{defn:ZAgeneralized}
For any $s\in\mathbb{R}$, determine ${\bf X}(z)={\bf X}(z;s)\in\mathbb{C}^{2\times 2}$ such that
\begin{enumerate}
	\item ${\bf X}(z)$ is analytic for $z\in\mathbb{C}\setminus\mathbb{R}$ and has a continuous extension on the closed upper and lower half-planes.
	\item The limiting values ${\bf X}_{\pm}(z)=\lim_{\epsilon\downarrow 0}{\bf X}(z\pm\I\epsilon),z\in\mathbb{R}$ satisfy the jump condition
	\begin{equation}\label{djump2}
		{\bf X}_+(z)={\bf X}_-(z)\begin{bmatrix}1-|r(z)|^2 & -\bar{r}(z)e^{-\I sz}\smallskip\\ r(z)e^{\I s z} & 1\end{bmatrix},\ \ z\in\mathbb{R}\ \ \ \ \ \textnormal{with}\ \ \ r(z)=-\I \int_\R \rmd t \, e^{-\I z t}A(t).
	\end{equation}
	\item As $z\rightarrow\infty$, we require the normalization
	\begin{equation}\label{eq:RiemannLaurentExp2}
		{\bf X}(z)=\mathbb{I}+{\bf X}_1z^{-1}+{\bf X}_2z^{-2}+\mathcal{O}\big(z^{-3}\big);\ \ \ \ \ {\bf X}_i={\bf X}_i(x,\gamma)=\big[X_i^{jk}(x,\gamma)\big]_{j,k=1}^2.
	\end{equation}
\end{enumerate}
\end{definition}
This Riemann-Hilbert problem is related to the Zakharov-Shabat system as
\begin{equation}
\label{eq:ZakharovShabatXgeneralized}
\frac{\partial {\bf X}}{\partial s}=\frac{\I z}{2} [{\bf X}, \sigma_3 ]+\I  \begin{bmatrix}
0 &q_0 \smallskip\\ 
-q_0 & 0 \end{bmatrix}{\bf X}
\end{equation}
We recast the above manipulations into a proposition providing the correspondence between the initial Fredholm determinant \eqref{eq:FredDet} and the Riemann-Hilbert problem associated to the Zakharov-Shabat system \eqref{eq:ZakharovShabatXgeneralized}.

\begin{proposition} \label{prop:ZSmoimoi}
The Fredholm determinant  $\Det(I-A_s^2)$ solves the Riemann-Hilbert problem associated to the Zakharov-Shabat in the sense that it generates the hierarchy of functions $\{ q_p, u_p\}_{p\in \N}$ providing the successive coefficients of the Laurent series as 
\begin{equation}\label{eq:explicitLaurentCoeffGeneralized}
{\bf X}_{p+1}=
\begin{bmatrix}
(-\I)^{p+1} u_p &\I^{p} q_p\smallskip\\ 
(-\I)^{p} q_p & \I^{p+1} u_p\end{bmatrix}
\end{equation}
and the explicit solution in terms of the operators $A_s$, $K_s=A_s^2$ and the resolvent of the derivative operator as
\begin{equation}\label{eq:explicitNewSolutionRiemannGeneralized}
{\bf X}(z)=\mathbb{I}+\bra{\delta}\begin{bmatrix}
-\frac{1}{D-\I z}A_s  &\frac{\I}{D+\I z} \smallskip\\ 
-\frac{\I}{D-\I z}  &- \frac{1}{D+\I z} A_s  \end{bmatrix}
 \frac{A_s}{I-K_s}\ket{\delta}.
\end{equation}

\end{proposition}
The main interpretation of Proposition~\ref{prop:ZSmoimoi} is the one-to-one correspondence between the determinantal point process with kernel $K_s$ and the Riemann-Hilbert problem of the Zakharov-Shabat system subject to the reflection coefficient which is the Fourier transform of the function $A$. The existence and uniqueness of both point process and solution to the Riemann-Hilbert problem seem therefore entangled. This leads us to formulate a very natural conjecture regarding the equivalence of the uniqueness condition in the two settings.

\begin{conjecture}[One uniqueness to rule them all]
The Riemann-Hilbert problem associated to the Zakharov-Shabat system with reflection coefficient $r(z)=-\I \int_\R \rmd u \, e^{\I z u}A(u)$ is uniquely solvable if and only if the determinantal point process with kernel $K_s=A_s^2$ is uniquely defined.
\end{conjecture}

Along with these results, we develop a few remarks on our solution. The first one being about the fact that the inverse of ${\bf X}$ verifies a system extremely close to the Zakharov-Shabat, indeed 
\begin{equation}
\label{eq:ZakharovShabatXgeneralizedInversed}
\frac{\partial {\bf X}^{-1}}{\partial s}=\frac{\I z}{2} [{\bf X}^{-1}, \sigma_3 ]-\I  {\bf X}^{-1} \begin{bmatrix}
0 &q_0 \smallskip\\ 
-q_0 & 0 \end{bmatrix}
\end{equation}

The second remark is the relation between the conservation laws expressed in Result.~\ref{result:FlowInvarianceI} and the Riemann-Hilbert problem \ref{defn:ZAgeneralized}.
From Eq.~\eqref{eq:ZakharovShabatXgeneralized}, it follows that the following trace is equal to zero for any $z\in \C$ and $s\in \R$
\begin{equation}
\label{eq:ConservationLawRHP}
\Tr\big({\bf X}^{-1}\frac{\partial {\bf X}}{\partial s}\big)=0
\end{equation}
Since we have the identity $\partial_s \Det({\bf X})=\Det({\bf X})\Tr\big({\bf X}^{-1}\frac{\partial {\bf X}}{\partial s}\big)$, it follows that the determinant of $ {\bf X}$ is independent of $s$. As we now show, all conserved quantities derive from this observation. Since Eq.~\eqref{eq:ConservationLawRHP} is true for any $z$, we can as well expand it as a Laurent series in $z$. To this aim, we require the exact expression of the matrix  ${\bf X}^{-1}$ which is obtainable by a simple $2\times 2$ matrix inversion as 
\begin{equation}
\Det({\bf X}) \times {\bf X}^{-1}=\mathbb{I}+\sum_{p=1}^\infty{\bf \tilde{X}}^{-1}_p z^{-p}, \qquad {\bf \tilde{X}}^{-1}_{p+1}=
\begin{bmatrix}
\I^{p+1} u_p &-\I^{p} q_p\smallskip\\ 
-(-\I)^{p} q_p & (-\I)^{p+1} u_p\end{bmatrix}
\end{equation}
Hence, the $p$-th coefficient of the Laurent series of \eqref{eq:ConservationLawRHP} is equal to zero leading to the identity
\begin{equation}
\label{eq:ConvationLawOrderP}
\partial_s \Tr ({\bf X}_p)=-\sum_{\ell=1}^{p-1}\Tr \left( {\bf \tilde{X}}^{-1}_\ell \partial_s {\bf X}_{p-\ell} \right) 
\end{equation}
Computing explicitly the matrix products and traces in Eq.~\eqref{eq:ConvationLawOrderP}, we obtain for odd $p$ the trivial identity $0=0$ and for even $p=2n+2$ we have 
\begin{equation}
 u'_{2n+1}= -\sum_{k=0}^{2n}(-1)^{k+1}\big[ u_{k}u'_{2n-k}-q_{k}q'_{2n-k} \big]
\end{equation}
which is exactly the flow invariance obtained in Result~\ref{result:FlowInvarianceI} upon integration. hence we have interpreted the flow invariants in the language of the Riemann-Hilbert problem.\\

We now raise a few questions and outlooks on our construction and result.
\begin{enumerate}
\item Since the Fredholm determinant $\Det(I-A_s^2)$ is numerically tractable by Bornemann's method, see \cite{bornemann2010numerical}, it provides a good indication as for the existence of an efficient numerical scheme for the Zakharov-Shabat system. 
\item We have seen in this work that determinants of the type $\Det(I-\sigma K_s)$ are related to integro-differential systems. It would be interesting to determine the generalization of the Zakharov-Shabat system taking for solution the Fredholm determinant constructed from a function $A$ and a measure $\sigma$.
\item The whole construction of this work has constrained the function $A$ to be real-valued, therefore we have not solved the Riemann-Hilbert problem for an arbitrary reflection coefficient whose Fourier transform is not real-valued. A promising direction to investigate more general reflection coefficients would be through Fredholm determinants built from a complex-valued function $A$ and a kernel $K=\bar{A}_sA_s$. The Fredholm determinant $\Det(I-\bar{A}_sA_s)$ would then generate a hierarchy with complex functions $\{ q_p, u_p \}$. 

\item The Zakharov-Shabat system is naturally related to the nonlinear Schrödinger equation \cite{shabat1972exact}. Indeed, introducing the differential equation $i\partial_t y=-\partial_{ss}^2 y +\frac{1}{2}\vert y \vert^2 y$, 
we can solve it by replacing the reflection coefficient in the Riemann-Hilbert problem \ref{defn:ZAgeneralized} as follows
\begin{equation}
r(z)\to r(z)e^{\I t z^2}
\end{equation}
so that the solution to the nonlinear Schrödinger equation - defocusing since the potential is repulsive in this case - is given by $y(s,t)=2\I X_1^{12}$.
Heuristically and conjecturally, it is tempting to define the Fredholm determinant $\Det(I-\bar{A}_{s,t} A_{s,t})$ with a time-dependent function $A_t$ given as 
\begin{equation}
r(z)e^{\I t z^2}=-\I \int_\R \rmd u \, e^{-\I z u}A_t(u).
\end{equation}
From the Fredholm determinant of kernel $\bar{A}_{s,t} A_{s,t}$, we expect to construct the first function of the hierarchy  $q_0=\bra{\delta} A_{s,t}(I-K_{s,t})^{-1}\ket{\delta}$  so that the solution at time $t$ reads $y(s,t)=2\I q_0$ or $y(s,t)=2\I \bar{q}_0$. Under mild conditions on the initial condition of the nonlinear Schrödinger equation, the mapping $y(s,t=0)\mapsto r(z)$, also called the direct scattering transform \cite{beals1984scattering}, is bijective. Hence, we can conjecture the following bijective procedure to solve the problem  
\begin{equation}
y(s,t=0) \to r(z) \to A_{s,t} \to y(s,t).
\end{equation}
Finally, since the focusing nonlinear Schrödinger equation is related to the Heisenberg spin chain model by the Hasimoto transform \cite{lakshmanan1977continuum}, it suggests that the Heisenberg chain can also be solved exactly through Fredholm determinants. We leave these fascinating questions for a future work. 
\item The above remark would provide a simple time and space dependence for the function $A_{s,t}$. Such dependence has also been observed in the exact solutions to the Kardar-Parisi-Zhang equation for a few cases, see Appendix~\ref{app:overview}, upon the replacement of the time exponentials $e^{\I t z^2} \to e^{\I t z^3}$. This might hint to the presence of a Zakharov-Shabat-like system in the framework of the KPZ equation and would then suggest a path to investigate whether its solution is always determinantal. 
\end{enumerate}

\section{Outlook and conclusion}

We have been interested throughout this work in presenting a framework gathering various problems that appeared at first disconnected: an integro-differential generalization of the Painlevé II hierarchy, some finite-time solutions to the Kardar-Parisi-Zhang equation, multi-critical fermions at finite temperature and the statistics of the largest real eigenvalue of the real Ginibre ensemble leading to a determinantal solution to the Zakharov-Shabat system. \\

The common ground was to identify a structure in the Fredholm determinants arising in these problems and to generalize a number of results that were previously obtained in the literature on an ad-hoc basis. Our main finding was a quasi-universal Hamiltonian system of equations for a hierarchy of functions and an infinite number of conserved quantities. We have shown in our setting that considering an inhomogeneous problem could be translated into integro-differential equations. We have presented for the first time a generalization of the second member of the Painlevé II hierarchy in terms of an integro-differential equation exactly solvable. Furthermore, we have extended the assertion that a notable Riemann-Hilbert problem related to the Zakharov-Shabat system could be solved in terms of Fredholm determinants. The natural sequel of that is the investigation of whether the solutions of the nonlinear Schrödinger equation can be obtained by Fredholm determinants explicitly.\\

Our work opens a few questions such as finding back some more features such as the quadratic invariant quantities and the hierarchy of column and matrix-like functions in the inhomogeneous Fredholm determinant setting by the means of the Riemann-Hilbert analysis, we believe that the construction of \cite{bothner2020origins} manipulating operator-valued Riemann-Hilbert problems would be the right starting point. Finally, the Fredholm determinants encountered in this work generally appear in the study of probabilistic systems \cite{krajenbrink2019beyond} and also in the context of quantum correlation in fermionic systems \cite{calabrese2010universal}. It would be interesting to obtain universal growth or large-deviation estimates directly from the infinite recursion and the conserved quantities unveiled. We leave all above questions open for further work.

\appendix

\section{Additional lemma and Pfaffian manipulations}
\label{app:pfaffOperator}
%
We first recall in this Appendix two useful lemma related to the differenciation of the resolvent of an operator with a kernel enjoying an additive structure and to the extension of the matrix determinant lemma in the case of a low-rank perturbation of a kernel. We additionally prove the Pfaffian manipulations of Results~\ref{result:SymplecPfaff} and \ref{result:OrthoPfaff}.

\begin{lemma}[Ferrari-Spohn derivative formula, {\cite[Lemma~3]{ferrari2005determinantal}}]
Let $A_s$ be an operator with a kernel with an additive structure, i.e. $A_s(x,y)=A(x+y+s)$, then the following holds
		\begin{equation}
		\partial_s\frac{I}{I+A_s}=\frac{A_s}{I-K_s}  D + \frac{A_s}{I-K_s}\ket{\delta} \bra{\delta} \frac{I}{I+A_s}
		\end{equation}
		where $D$ is the derivative operator.
		\label{lemma:FerrariSpohn}
\end{lemma}

\begin{lemma}[Extended matrix determinant lemma]
\label{lemma:MatDetLemEx}
Given a kernel $K_s$ perturbed by a sum of $m$ rank-one operators $\ket{f_i}\bra{g_i}$ of kernel $(x,y)\mapsto f_i(x)g_i(y)$ for $i=1,\dots,m$, we have 
\begin{equation}
\Det(I-K_s-\sum_{i=1}^m \ket{f_i}\bra{g_i})=\Det(I-K_s)\Det\big[ \delta_{ij}-\bra{f_i}\frac{I}{I-K_s}\ket{g_j}\big]_{i,j=1}^m
\end{equation}
where $\delta_{ij}$ is the Kronecker's delta.
\end{lemma}

We turn to the proofs of Results~\ref{result:SymplecPfaff} and \ref{result:OrthoPfaff} to relate the symplectic/orthogonal determinants to their Pfaffian counterpart. The proofs are mostly technical so we detail the few central points before entering the core of the computation and we refer to Ref.~\cite{tracy1996orthogonal} for their first appearance in the literature. \\

Starting from the Fredholm Pfaffian represenration, the first main idea is to use a factorization identity of the type
\begin{equation}
\mathrm{Pf}(J-K)=\mathrm{Pf}( J-A^{(1)}A^{(2)})
\end{equation}
Using that for a skew-symmetric kernel $K$, $\mathrm{Pf}[J-K]^2=\Det[I+JK]$,  where the scalar kernel $I$ is the identity kernel, this gives
\begin{equation}
\mathrm{Pf}( J-K)^2=\Det( I+JA^{(1)}A^{(2)})
\end{equation}
Following \cite{tracy1998correlation}, using Sylvester's identity $\Det(I+AB)=\Det(I+BA)$ for arbitrary Hilbert-Schmidt operators $A$ and $B$. They may act between different spaces as long as the products make sense. In the present context $\Det(I+AB)$ is the Fredholm determinant of a matrix-valued kernel whilst $\Det(I+BA)$ is a Fredholm determinant of scalar-valued kernel. Hence we obtain
\begin{equation}
\mathrm{Pf}( J-K) ^2=\Det (I+A^{(2)}JA^{(1)})
\end{equation}
Further manipulations on the scalar kernel will be necessary to conclude. Let us now start with the symplectic case. \\

\begin{proof}[Proof of the symplectic determinant Pfaffian relation, Result~\ref{result:SymplecPfaff}]
We define the antisymmetric operator $B^{(\rm sympl)}: \mathbb{L}^2(\R_+) \to \mathbb{L}^2(\R_+) $ depending on the operators $A_s$ and $K_s$ as 
\begin{equation}
\begin{split}
B^{(\rm sympl)}=D^{-1}K_s+\frac{1}{2}A_s\ket{1}\bra{1}A_s.
\end{split}
\end{equation}
The kernel of the operator vanishes exponentially for both variables at $+\infty$, hence all integration by parts will only the values at 0. For readability purpose, we will subsequently omit the subscript \textit{"sympl"} for the operator. The first steps of the computation which lead in the factorization read
\begin{equation}
\begin{split}
{\rm Pf} \left(J-
\begin{bmatrix}
B &-BD^\intercal\\ 
-DB &  DBD^\intercal\end{bmatrix}\right)^2
&={\rm Pf} \left(J-
\begin{bmatrix}
1 &0\\ 
0 &  D\end{bmatrix}
\begin{bmatrix}
B &-BD^\intercal\\ 
-B &  BD^\intercal\end{bmatrix}\right)^2\\
&=\Det\left( I + \begin{bmatrix}
B &-BD^\intercal\\ 
-B &  BD^\intercal\end{bmatrix} J \begin{bmatrix}
1 &0\\ 
0 &  D\end{bmatrix}
\right)\\
&=\Det\left( I + \begin{bmatrix}
BD^\intercal &BD\\ 
 -BD^\intercal & - BD\end{bmatrix} 
\right)\\
\end{split}
\end{equation}
Summing the first line to the second one and subtracting the second column to the first one, we obtain 
\begin{equation}
\begin{split}
=\Det\left( I + \begin{bmatrix}
BD^\intercal &BD\\ 
 -BD^\intercal & - BD\end{bmatrix} 
\right)
&=\Det\left( I + \begin{bmatrix}
B(D^\intercal -D)&BD\\ 
0 & 0\end{bmatrix} 
\right)\\
&=\Det\left(I+B(D^\intercal-D)\right)\\
&=\Det\left(I+2BD^\intercal+B\ket{\delta}\bra{\delta}\right)\\
&=\Det\left( I-2DB \right)\big[ 1+\bra{\delta} \frac{I}{I+2BD^\intercal} B \ket{\delta} \big]\\
\end{split}
\end{equation}
From the first line to the second line, we used a block determinant identity to reduce the matrix kernel into a scalar one. From the second to the third line, we used the integration by parts identity $D^\intercal +D=-\ket{\delta}\bra{\delta}$ valid since the kernel vanishes towards $+\infty$. From the third line to the fourth one we used the matrix determinant lemma \ref{lemma:MatDetLem} allowing to evaluate rank-one operator contributions in a Fredholm determinant and transposed the operator $BD^\intercal$ in the remaining Fredholm determinant using $B^\intercal=-B$. We will now show that the last term is equal to zero
\begin{equation}
Q\equiv 2 \bra{\delta} \frac{I}{I+2BD^\intercal} B \ket{\delta}=0
\end{equation}
This was already proved in Ref.~\cite[Appendix~B]{krajenbrink2018large} and we recall the computation for completeness. The main arguments are the antisymmetry of $B$, the integration by parts identity $D^\intercal=-D-\ket{\delta}\bra{\delta}$ and the  commutation relation $(I+BD^\intercal )^{-1}B=B(I+D^\intercal B)^{-1}$.
Taking the adjoint of the operator $(I+2BD^\intercal)^{-1}(2B)$, we have
\begin{equation}
\begin{split}
Q&=-\bra{\delta} 2B  \frac{I}{I-2DB}\ket{\delta}=-\bra{\delta} 2B\frac{I}{I+2D^\intercal B+2\ket{\delta}\bra{\delta}B} \ket{\delta}
\end{split}
\end{equation}
Using the Sherman-Morrison identity as the last term in the inverse is a rank-one operator, we obtain
\begin{equation}
\begin{split}
Q&=-2 \bra{\delta} B\frac{I}{I+2D^\intercal B}  \ket{\delta}+4\frac{\bra{\delta} B\left(I+2D^\intercal B \right)^{-1}\ket{\delta}\bra{\delta} B\left(I+2D^\intercal B \right)^{-1}k\ket{\delta}}{1+2\bra{\delta}B\left(I+2D^\intercal B\right)^{-1}\ket{\delta}}\\
&=-Q+\frac{Q^2}{1+Q}
\end{split}
\end{equation}
which implies $Q=0$ or $Q=-2$. Since the amplitude of $B$ can be increased continuously from $0$ to any value, by continuity, the solution is $Q=0$.
We finally define $K^{(\rm sympl)}=2DB^{(\rm sympl)}$ to conclude the derivation.

\end{proof}

We now turn to the orthogonal case. \\

\begin{proof}[Proof of the orthogonal determinant Pfaffian relation, Result~\ref{result:OrthoPfaff}]
We define the antisymmetric operator $B^{(\rm ortho)}: \mathbb{L}^2(\R_+) \to \mathbb{L}^2(\R_+) $ depending on the operators $A_s$ and $K_s$ as 
\begin{equation}
B^{(\rm ortho)}=D^{-1}K_s+\frac{1}{2}A_s\ket{1}\bra{1}A_s +\frac{1}{2}\big( \ket{1}\bra{1}A_s-A_s\ket{1}\bra{1}\big)
\end{equation}
The operator of this kernel does not vanish towards $+\infty$ and therefore all integration by parts will have to include the contribution at 0. To this aim, we introduce for this proof the two notations $\ket{\delta_0}$ as the projector to 0 and $\ket{\delta_\infty}$ as the projector to $+\infty$ so that
\begin{equation}
D+D^\intercal= \ket{\delta_\infty}\bra{\delta_\infty}-\ket{\delta_0}\bra{\delta_0}
\end{equation}
We also omit the subscript \textit{"ortho"} in the subsequent computations. As for the symplectic case, the first steps of the derivation read 

\begin{equation}
\begin{split}
{\rm Pf} \left(J-
\begin{bmatrix}
B-\varepsilon &-AB^\intercal\\ 
-DB &  DBD^\intercal  \end{bmatrix}\right)^2&={\rm Pf} \left(J-
\begin{bmatrix}
1 &0\\ 
0 &  D\end{bmatrix}
\begin{bmatrix}
B-\varepsilon&-BD^\intercal\\ 
-B &  BD^\intercal \end{bmatrix}\right)^2\\
&=\Det\left( I + \begin{bmatrix}
B-\varepsilon &-BD^\intercal\\ 
-B & BD^\intercal\end{bmatrix} J \begin{bmatrix}
1 &0\\ 
0 &  D\end{bmatrix}
\right)\\
&=\Det\left( I + \begin{bmatrix}
BD^\intercal &BD-\varepsilon D\\ 
-BD^\intercal & -BD\end{bmatrix} 
\right)
\end{split}
\end{equation}
Summing the second line to the first one and subtracting the first column to the second one,  we obtain  

\begin{equation}
\begin{split}
\Det\left( I + \begin{bmatrix}
BD^\intercal &BD-\varepsilon D\\ 
-BD^\intercal & -BD\end{bmatrix} 
\right)
&=\Det\left( I + \begin{bmatrix}
0&-\varepsilon D\\ 
-BD^\intercal & B(D^\intercal -D)\end{bmatrix} 
\right)\\
&=\Det\left(I-BD+BD^\intercal(I-\varepsilon D)\right)\\
&=\Det\left(I-DB+D^\intercal(I-\varepsilon D)B\right)
\end{split}
\end{equation}
To go from the first line one with a matrix-valued kernel to the second line with a scalar kernel, we used a block determinant formula, to go from the second line to the third one we used Sylverster's identity to commute the position of $B$. Recalling that the operator $\varepsilon$ is the sign operator verifying
\begin{equation}
D\varepsilon=I, \qquad \varepsilon\ket{\delta_0}=\frac{1}{2}\ket{1}, \qquad \varepsilon\ket{\delta_\infty}=-\frac{1}{2}\ket{1},
\end{equation}
the integration by parts on $\varepsilon$ (which does not vanish at $+\infty$) reads 
\begin{equation}
\begin{split}
\varepsilon D&=-\varepsilon D^\intercal - \varepsilon \big[\ket{\delta_0}\bra{\delta_0}-\ket{\delta_\infty}\bra{\delta_\infty}\big]\\
&=I - \frac{1}{2}\varepsilon \ket{1}\big[\bra{\delta_0}+\bra{\delta_\infty}\big]\\
&=I - \frac{1}{2}\varepsilon \ket{1}\big[2\bra{\delta_\infty}-\bra{1}D\big]
\end{split}
\end{equation}
where we have used from the second line to the third one that $\bra{1}D=\bra{\delta_\infty}-\bra{\delta_0}$. Hence
we have
\begin{equation}
\label{eq:PfaffIntermediateCalc}
\begin{split}
\Det\left(I-DB+D^\intercal(I-\varepsilon D)B\right)
&=\Det\left(I-DB+\frac{1}{2}D^\intercal \ket{1}\big[2\bra{\delta_\infty}-\bra{1}D\big]B\right)
\end{split}
\end{equation}
Let's have a look at the contribution of the second part of the projector. By the determinant lemma, it reads
\begin{equation}
\bra{1} DB \frac{I}{I-DB}D^\intercal \ket{1}=-\bra{1}D^{\intercal}\ket{1}+\bra{1}  \frac{I}{I-DB}D^\intercal \ket{1}
\end{equation}
Since the inner product $\bra{1}D^{\intercal}\ket{1}$ is equal to 0, we simplify Eq.~\eqref{eq:PfaffIntermediateCalc} by modifying the projector and further calculate the action of $\bra{\delta_\infty}$ on $B$ as
\begin{equation}
\bra{\delta_\infty}B=\frac{1}{2}\bra{1}A_s, \qquad B\ket{\delta_\infty}=-\frac{1}{2}A_s\ket{1}
\end{equation}
to obtain 
\begin{equation}
\begin{split}
\Det\left(I-DB+\frac{1}{2}D^\intercal \ket{1}\big[2\bra{\delta_\infty}-\bra{1}D\big]B\right)
&=\Det\left(I-DB-\frac{1}{2}D^\intercal \ket{1}\bra{1}(I-A_s)\right)\\
&=\Det\left(I-A_s^2-\frac{1}{2}(I-A_s)D^\intercal\ket{1}\bra{1}(I-A_s) \right)\\
\end{split}
\end{equation}
From the first to the second line, we replaced the exact expression of $B^{(\rm ortho)}$ as a function of $A_s$. We now investigate the projector once again through the matrix determinant lemma 
\begin{equation}
\begin{split}
\Det\left(I-A_s^2-\frac{1}{2}(I-A_s)D^\intercal\ket{1}\bra{1}(I-A_s) \right)\
&= \Det\left(I-A_s^2 \right) \left(1-\frac{1}{2}\bra{1}\frac{I-A_s}{I+A_s}D^\intercal\ket{1}\right)\\
&= \Det\left(I-A_s^2 \right) \left(1+\bra{1}\frac{A_s}{I+A_s}D^\intercal\ket{1}\right)\\
&= \Det\left(I-A_s^2 \right) \left(1-\bra{1}\frac{A_s}{I+A_s}\ket{\delta_0}\right)\\
&= \Det\left(I-A_s \right)\Det\left(I+A_s-\ket{\delta_0}\bra{1}A_s \right) \\
&= \Det\left(I-K_s-A_s\ket{\delta_0}\bra{1}(I-A_s) \right)\\
\end{split}
\end{equation}
To go from the second to the third line, we used that $D^\intercal \ket{1}=-\ket{\delta_0}$ since $A_s$ vanishes at $+\infty$. From the third line to the fourth one, we separated $\Det(I-A_s^2)$ into $\Det(I+A_s)\Det(I-A_s)$ and used the Sherman-Morrison lemma to reinject the projector in the determinant. We finally define $K^{(\rm ortho)}=K_s+A_s\ket{\delta_0}\bra{1}(I-A_s) $ to conclude the derivation.

\end{proof}

For completeness, we recall another proposition which states the equivalence between a Fredholm Pfaffian of symplectic type with another one involving the $\delta'$ operator. 
~\begin{proposition}[Matrix kernel equivalence, {\cite[Proposition 5.2]{baik2018pfaffian}}]\label{prop:barraquand}
Let $\mathbf{B}:\R^2 \to \rm{Skew}_2(\R)$ be a kernel of the form
	$$ \mathbf{B}(x,y) =  \begin{pmatrix}
	B(x,y) & -\partial_y B(x, y) \\
	-\partial_x B(x,y) & \partial_x\partial_y B(x,y)
	\end{pmatrix},$$
	where $B$ is smooth, antisymmetric, and $\mathbf{B}$ satisfies the following decay hypotheses:
there exist constants $c>0$ and $a>b\geqslant 0$ such that
	\begin{equation*}
	\vert B(x, y) \vert <c e^{-ax-ay}, \ \
	\vert \partial_yB(x, y)\vert   <c e^{-ax+by}, \ \
	\vert \partial_x \partial_yB(x, y)\vert  <c e^{bx+by}.
	\end{equation*}	
	Let $\mathbf{C}$ be the kernel
	$$ \mathbf{C}(x,y) =  \begin{pmatrix}
	B(x,y) & -2\partial_y B(x, y) \\
	-2\partial_x B(x,y) & 4\partial_x\partial_y B(x,y) +\delta'(x,y)
	\end{pmatrix}.$$
	where $\delta'$ is a distribution on $\R^2$ such that
\begin{equation} \iint f(x,y)\delta'(x,y)\mathrm{d}x\mathrm{d}y =  \int  \left.\big(\partial_y f(x,y) - \partial_x f(x,y)\big)\right|_{y=x}\mathrm{d}x,\label{eq:defdeltaprime}
\end{equation}
for smooth and compactly supported test functions $f$. 	Then for any $s\in \R$,
	\begin{equation}
	\ensuremath{\mathrm{Pf}}[J-\mathbf{B}]_{\mathbb{L}^2(s, +\infty)}= \ensuremath{\mathrm{Pf}}[J-\mathbf{C}]_{\mathbb{L}^2(s, +\infty)}.
	\label{eq:equivalentpfaffians}
	\end{equation}
\end{proposition}


\section{Overview of the existing kernels in the literature}\label{app:overview}
We recall in this Section a few examples of situations where Fredholm determinants of the structure of \eqref{eq:FredDet} and \eqref{eq:FredDet1} appear in the literature. We will be specifically inspired by the literature of random matrix theory and of exact solutions to the Kardar-Parisi-Zhang equation in 1+1 dimensions.

\subsection{Largest eigenvalue of Gaussian random matrices}\label{subsec:appGaussian}

Our first focus is directed towards the Gaussian $\beta$ ensemble (G$\beta$E) of random matrices \cite{mehta2004random}. To this aim, we consider a matrix of size $N\times N$ such that its eigenvalues are real, labeled $\lbrace \lambda_i \rbrace$,  with a joint probability distribution function (JPDF) of the form (up to a normalization constant)
\be \label{PDF}
P[\lbrace \lambda_i \rbrace] \propto  \exp \left( \beta \sum_{1 \leqslant i < j \leqslant N} \log |\lambda_i-\lambda_j|
- \frac{\beta N}{4} \sum_{i=1}^N \lambda_i^2 \right).
\ee
The Gaussian weight in Eq.~\eqref{PDF} is the reason for the name of the Gaussian ensemble. The logarithmic contribution of the JPDF also takes the form of a Vandermonde factor $\prod_{i<j} \abs{\lambda_i-\lambda_j}^\beta$, indicating a strong correlation between the eigenvalues.  With these conventions, in the large $N$ limit the empirical measure $ \Lambda_N(\lambda) := N^{-1} \sum_{i=1}^N \delta_{\lambda_i}(\lambda) $ converges to the celebrated Wigner semi-circle distribution \cite{mehta2004random} with density $\Lambda_\text{sc}(\lambda) = \frac{1}{2\pi}\sqrt{4-\lambda^2}\ind_{\lbrace |\lambda|<2 \rbrace }$.\\

Historically, matrix representations for the G$\beta$E were obtained solely for $\beta=1,2,4$ and are referred to as the orthogonal (GOE), unitary (GUE) and symplectic (GSE) ensembles (due to a conjugacy symmetry of the matrix leaving the spectrum invariant) \cite{mehta2004random}. For completeness, let us describe the construction of a GOE matrix. Denote $M$ the $N\times N$ real symmetric matrix whose entries above the diagonal are independent random Gaussian variables with mean zero and variance
\begin{equation}
\mathbb{E}\left[ M_{ij}^2\right]=\frac{1+\delta_{ij}}{N}.
\end{equation}
Then the probability measure of $M$ is given, up to normalization, by 
\begin{equation}
P(M)\propto \exp\big(-\frac{N}{4}\Tr M^2\big)
\end{equation}
and the JPDF of its eigenvalues is given up to normalization by Eq.~\eqref{PDF}. 

 Due to the presence of the $\beta$ factor, the JPDF \eqref{PDF} can be seen as the Gibbs measure 
of a Coulomb gas (CG) with logarithmic repulsion between the eigenvalues, which, at large $N$, are described by a continuous density. In addition, this JPDF is the stationary measure of the $\beta$ Dyson Brownian motion \cite{dyson1962brownian} which represents particles $\lbrace \lambda_i \rbrace $ driven by a Brownian motion and interacting with a logarithmic potential $\log \abs{\lambda_i-\lambda_j}$. In addition, in the case of $\beta=2$, if we see the eigenvalues $\lbrace \lambda_i \rbrace$ as the positions of identical non-interacting fermionic particles as in Ref.~\cite{dean2015finite,dean2016noninteracting}, then the JPDF can be represented as the square modulus of the related $N$-body fermionic wave function, i.e. the quantum probability.\\

A particular feature of the random eigenvalues is their behavior around the edge of the spectrum located at $\lambda=2$ for large $N$. Near the edge, the fluctuations of the eigenvalues are stronger than in the bulk of the spectrum and a non-trivial behavior is found in a window of width
$\sim N^{-2/3}$ around the edge. In that window for large $N$, the scaled eigenvalues $a_i \equiv N^{2/3} (\lambda_i-2)$  define the Airy$_\beta$ point process which describes the few largest eigenvalues of a large G$\beta$E matrix.\\

 In the case of $\beta=2$, the Airy$_2$ point process has the special structure of a determinantal point process. It is an infinite random point configuration ${\bf a}=(a_1 > a_2 > \cdots)$ on $\R$. Its mean density $\rho(a)$ (seen as the average of the empirical density of $a$) is equal to $ \rho(a)=K_\Ai(a,a)$ 
with the Airy kernel $K_\Ai$. More generally, the $k$-th correlation function $\rho_k(x_1,\dots,x_k)$ for all $k\geqslant 1$ takes a determinantal form
\begin{equation}
\rho_k(x_1,\dots,x_k)=\det[K_\Ai(x_i,x_k)]_{i,j=1}^k.
\end{equation}

For $\beta=1,2,4$, the cumulative distribution of the largest Airy point $a_1$ is the celebrated Tracy-Widom (TW) distribution \cite{tracy1994level, tracy1996orthogonal, tracy1998correlation}. Equivalently, denoting $\lambda_{\max}$ the largest eigenvalue of a G$\beta$E matrix, we have
\begin{equation}
\lim_{N\to +\infty} \mathbb{P}\big(\frac{\lambda_{\max}-2}{N^{-2/3}}\leqslant s\big)= \mathbb{P}(a_1\leqslant s):=F_\beta(s)
\end{equation}

\begin{figure}[t!] 
\begin{center}
\includegraphics[scale=0.7]{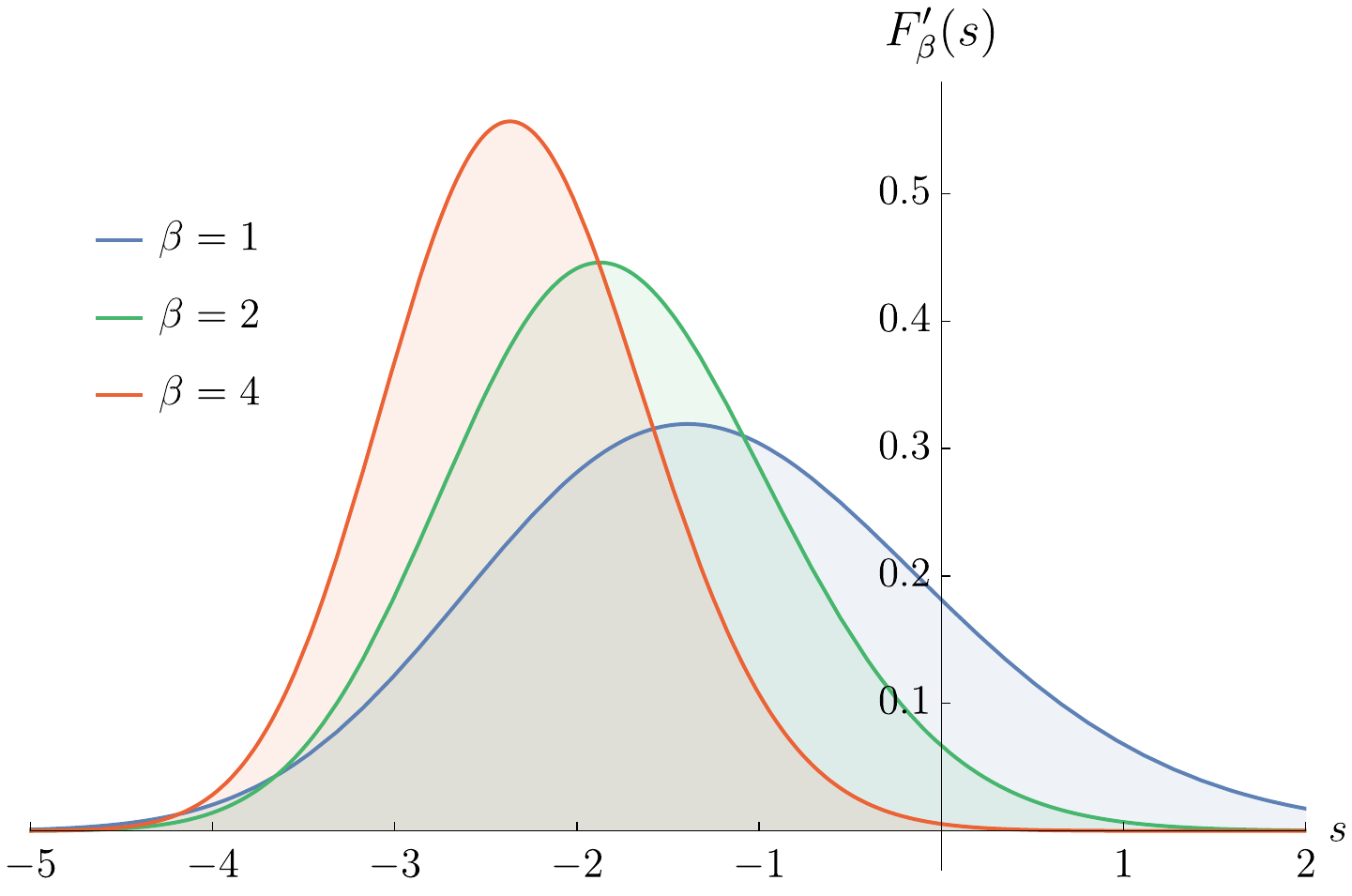}
\caption{Tracy-Widom distributions for the GOE, $\beta=1$ (blue line), the GUE, $\beta=2$ (green line) and the GSE, $\beta=4$ (red line). The plots were performed on Mathematica with the dedicated Tracy-Widom distribution function.}
\label{fig:TWPDFplot}
\end{center}
\end{figure}

The Tracy-Widom distributions for $\beta=1,2,4$ are intrinsically related to Fredholm determinants of the Airy kernel. Indeed, for $s$ in $\R$, let $\Ai_s$ be the Airy integral operator constructed from the kernel $\Ai_s(x,y)=\Ai(x+y+s)$ 
and $K_{\Ai,s}$ the shifted Airy operator with kernel $K_{\Ai,s}(x,y)=\int_0^{+\infty} \rmd r \, \Ai(x+r+s)\Ai(y+r+s)$ where $\Ai$ is the standard Airy function. The Tracy-Widom distributions for $\beta=1,2,4$ admit the representations
\begin{itemize}
\item $F_2(s)=\Det(I- K_{\Ai,s})$
\item $F_1(s)=\sqrt{\Det(I-K_{\rm Forr})}=\Det(I- \Ai_s)$
\item $F_4(s)=\sqrt{\Det(I- K^{\rm GLD})}=\frac{1}{2}\left(\Det(I- \Ai_s)+\Det(I+ \Ai_s)\right)$
\end{itemize}
All operators are considered on $\mathbb{L}^2(\R_+)$ and the intermediate kernels $K_{\rm Forr}$  and $K_{\rm GLD}$ are given by
\begin{equation}
\begin{split}
K_{\rm Forr}(x,y)&=K_\Ai(x+s,y+s)-\Ai(x+s)\left(1-\int_0^{+\infty} \rmd \lambda \, \Ai(s+y+\lambda)\right),\\
K^{\rm GLD}(x,y)&=K_{\rm Ai}(s+x,s+y)-\frac{1}{2}{\rm Ai}(s+x) \int_0^{+\infty} \rmd \lambda \, \Ai(s+y+\lambda).
\end{split}
\end{equation}

The kernel $K_{\rm Forr}$ was shown by Forrester in Ref.~\cite{forrester2000painlev}  to be related to the GOE Tracy-Widom distribution function $F_1(s)$, the kernel $\Ai_s$ was proved to be related to $F_1(s)$ by Ferrari and Spohn in Ref.~\cite{ferrari2005determinantal} and the kernel $K^{\rm GLD}$ was shown by Gueudré and Le Doussal in Ref.~\cite{gueudre2012directed} to be related to the GSE Tracy-Widom distribution function $F_4(s)$. The kernels $K_{\rm Forr}$ and $K^{\rm GLD}$ are the orthogonal and symplectic-like kernels introduced in \ref{subsec:OrthoSymplecLike} as a rank-one perturbation of the Airy kernel.
\begin{remark}
Here $F_4(s)$ is the cumulative distribution function of the GSE-TW distribution, as defined in \cite{baik2008asymptotics}. 
Another convention, which we denote $\tilde F_4$ with $F_4(s)=\tilde F_4(\frac{s}{\sqrt{2}})$,
is given in \cite{tracy1996orthogonal,ferrari2005determinantal}. 
\end{remark}

Even though the Tracy-Widom distributions are expressed through Fredholm determinants of integral operators, there exist very efficient numerical schemes to compute them due to Bornemann, see Refs.~\cite{bornemann2010numerical,bornemann2011accuracy}.

\subsection{Correlation functions of the elliptic Ginibre ensembles}
\label{subsec:appEllipticGinibre}
\begin{figure}[t!] 
\begin{center}
\includegraphics[scale=0.65]{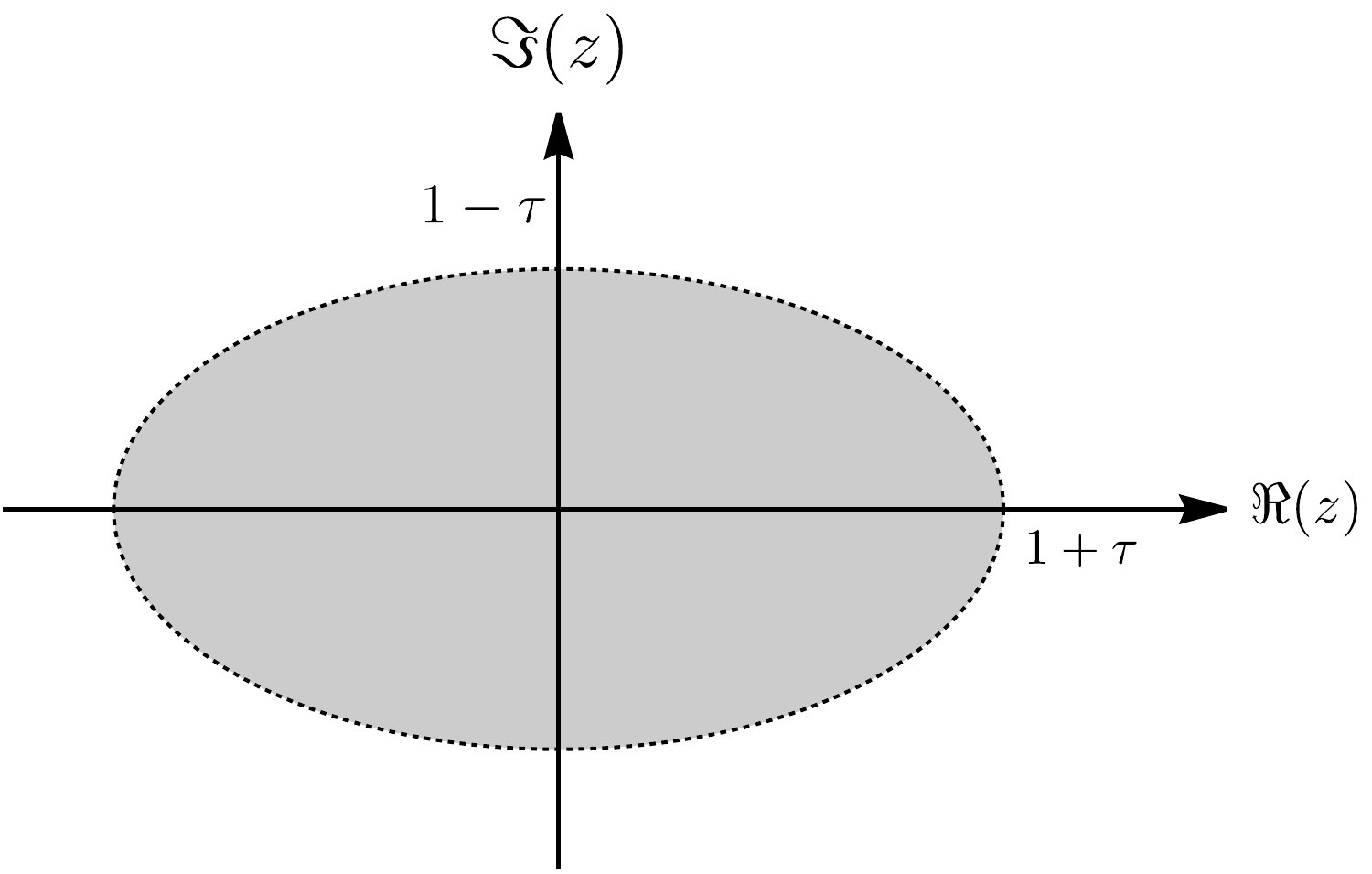}
\caption{Representation of the elliptic law. The density of eigenvalues in the complex plane is uniform within the ellipse of half-axes of length $1+\tau$ and $1-\tau$ interpolating between the circular law of the Ginibre ensemble and the semi-circular law of the Gaussian ensemble.}
\label{fig:Elliptic}
\end{center}
\end{figure}

This Appendix is inspired from the presentation of Elliptic ensembles of Refs.~\cite{bender2010edge, akemann2012universality}. 
The elliptic Ginibre ensembles, denoted similarly to the Gaussian ensembles, GinOE, GinUE and GinSE are a family of random matrices of size $N\times N$, depending on a parameter $\tau\in [0,1)$, defined by the probability measure for a matrix $M$, up to normalization
\begin{equation}
P(M)\propto \exp\big( -N \frac{\gamma_\beta  }{1-\tau^2}\Tr(MM^\dagger -\frac{\tau}{2}(M^2 +M^{\dagger 2})) \big) 
\end{equation}
The parameters read $\gamma_{\beta=2}=1$ and $\gamma_{\beta=1,4}=\frac{1}{2}$.
As an example of concrete realization of this ensemble for $\beta=2$, the complex case, take two independent GUE matrices $H_1$ and $H_2$ such that
\begin{equation}
M_{\beta=2}=\sqrt{\frac{1+\tau}{2}}H_1+\I \sqrt{\frac{1-\tau}{2}}H_2
\end{equation} the probability distribution reads
\begin{equation}
P(M)\propto \exp\big( -\frac{\gamma_\beta N }{1+\tau}\Tr(H_1^2)-\frac{\gamma_\beta N }{1-\tau}\Tr(H_2^2)\big) 
\end{equation}
we observe that
\begin{itemize}
\item for $\tau=0$, the distribution factorizes between the Hermitian and anti-Hermitian part and there is no particular symmetry: this is the usual Ginibre ensemble;
\item for $\tau\to 1$, the distribution converges to the one of the usual Gaussian $\beta$ ensemble and the eigenvalues become all real.
\end{itemize}
In the large $N$ limit, the empirical measure of the eigenvalues $\{ z_j \}_{j=1,\dots,N}$ converges to a generalization of the Wigner semi-circle called the elliptic law. The limited density is a uniform measure
\begin{equation}
\Lambda_{\rm ell}(z)=\frac{1}{\pi(1-\tau^2)}\mathds{1}_{z\in \rm{ell}}
\end{equation}
within the ellipse of half-axes of length $(1+\tau)$ and $(1-\tau)$ represented in Fig.~\ref{fig:Elliptic}. \\

Focusing on $\beta=2$, it is known that the random eigenvalues form a determinantal form process \cite{bender2010edge, akemann2012universality}. In the large $N$ limit, taking the cross-over parameter for the almost Hermitian regime $\sigma= N^{1/6} \sqrt{1-\tau}$, the eigenvalues fluctuate close to the real axis and scale as 
\begin{equation}
z=(1+\tau) +\frac{x+\I y}{N^{2/3}}
\end{equation}
The cross-over kernel of the determinantal point process yields in this regime (with $z=x+\I y$)
\begin{equation}
K(z_1,z_2)=\frac{1}{\sigma\sqrt{\pi}}e^{-\frac{y_1^2+y_2^2}{2\sigma^2}+\frac{\sigma^6}{6}+\frac{\sigma^2(z_1+\bar{z}_2)}{2}}\int_{\R_+}\rmd r \, e^{\sigma^2 r}\Ai(z_1+r+\frac{\sigma^4}{4})\Ai(\bar{z}_2+r+\frac{\sigma^4}{4})
\end{equation}
In the Hermitian limit $\sigma\to 0$, we have $K(z_1,z_2)\to \sqrt{\delta(y_1)\delta(y_2)}K_{\Ai}(x_1,x_2)$. This kernel has the same multiplicative structure as in \eqref{eq:KernelStructureMul} and closely resembles \eqref{Ktau} upon the projection onto $y_1,y_2=0$.
\subsection{Solutions at all times to the Kardar-Parisi-Zhang equation}
\label{recall}
Another situation where the Fredholm determinants of interest appear is the study of the exact solutions to the Kardar-Parisi-Zhang (KPZ) equation in 1+1 dimensions \cite{KPZ}. Consider the KPZ equation on the real line for the height field $h(x,t)$
\begin{equation}
\partial_t h = \partial_{xx}^2 h+(\partial_x h)^2 +\sqrt{2}\eta(x,t)
\end{equation}
where $\eta(x,t)$ is a space-time white noise with correlator $\E[\eta(x,t)\eta(x',t')]=\delta(x-x')\delta(t-t')$, we will focus on two particular initial geometries called the droplet and the Brownian initial conditions. More specifically we will discuss the generating function of the exponential of the  KPZ height which exhibits a Fredholm determinant representation for the aforementioned geometries. 

\subsubsection{Droplet initial condition}
The droplet initial condition is defined as
\begin{equation}
h(x,t=0)=-w\abs{x}+\log(w/2), \quad w\gg 1
\end{equation} 
and we will be interested  in the  moment generating function of the exponential shifted height 
\begin{equation}
H(t)=h(0,t)+\frac{t}{12}.
\end{equation}
The solution of the KPZ equation for the droplet initial condition was found originally by several groups and was presented in Refs.~\cite{sasamoto2010one, ACQ, dotsenko,CalabreseDR} . The moment generating function is given in terms of a Fredholm determinant with the Airy kernel and the so-called Fermi factor measure. 
\begin{equation}
\label{eq:G[M] sw}
 \mathbb{E}_{\mathrm{KPZ}}\left[ \exp\left( -z e^{H(t)} \right)\right]=\mathrm{Det}(I-\sigma_{z,t}K_\Ai)_{\mathbb{L}^2(\mathbb{R})}\;.
\end{equation}
where the expectation value is taken over the white noise of the KPZ equation, $K_\Ai$ is the Airy kernel, $K_{\Ai}(u,u')=\int_{0}^{\infty} \! \rmd r \; \Ai(r+u) \Ai(r+u')$, 
and the weight of the Airy kernel $\sigma_{z,t}$ is the Fermi factor expressed as
\begin{equation} \label{M sw} 
\sigma_{z,t}(u) = \frac{z}{ z+  e^{- t^{1/3} u}}
\end{equation}
Defining $z=e^{-st^{1/3}}$ allows to fit the determinant \eqref{eq:G[M] sw} in our framework. At late time, the Fermi factor becomes a projector: $\sigma_{z,t}(u)\to_{t\to +\infty} \Theta(u-s)$, the double exponential becomes an indicator function $e^{-e^{\lambda (\cdot)}}\to_{\lambda \to +\infty}\mathds{1}(\cdot \leqslant 0)$ and hence the cumulative distribution of the shifted height $H(t)=h(0,t)+\frac{t}{12}$ converges to the Tracy-Widom distribution for $\beta=2$.
\begin{equation}
\lim_{t \to +\infty}\mathbb{P}\big(\frac{H(t)}{t^{1/3}}\leqslant s\big)=F_2(s) 
\end{equation}

\subsubsection{Brownian initial condition}
Another initial condition that has been solved is the Brownian one, sometimes also called the stationary interface in the case of zero drift. Mathematically, it is described by a two-sided Brownian interface pinned at $x = 0$
\begin{figure}[t!] 
\begin{center}
\includegraphics[scale=0.6]{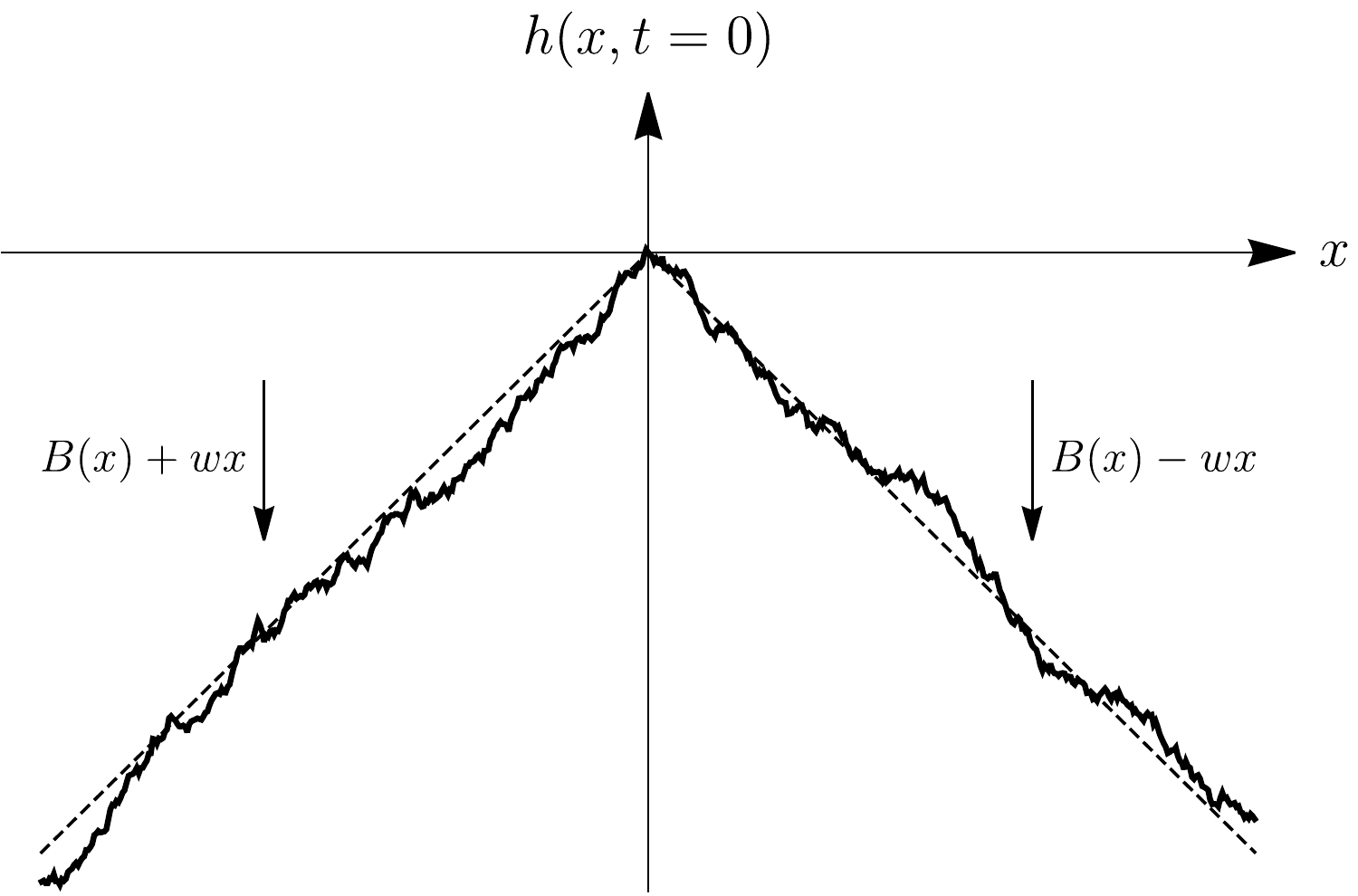}
\caption{Two-sided Brownian motion with drift $w$ as the initial condition to the Kardar-Parisi-Zhang equation.}
\label{fig:Brownian}
\end{center}
\end{figure}
\begin{equation}
h(x,t=0)=-w\abs{x}+B(x)
\end{equation}
where $B(x)$ is a standard two-sided Brownian motion. We represent this initial condition in Fig.~\ref{fig:Brownian}. In addition to averaging over realizations of the white noise of the KPZ equation, one has to average over all possible initial Brownian interfaces to obtain the moment generating function of the KPZ height. Imamura and Sasamoto \cite{imamura2012exact} and  Borodin, Corwin, Ferrari, Veto \cite{borodin2015height} derived the exact explicit representations for the generating function in terms of a Fredholm determinant and an additional random variable $\chi$ independent of $h(x,t)$, with probability density $p(\chi)\rmd \chi =   e^{-2 w \chi -  e^{-\chi}} / \Gamma(2 w)\rmd \chi $.

\begin{equation} \label{QGamma} 
 \mathbb{E}_{\mathrm{KPZ}, B}\left[ \exp\left( -z e^{\chi+h(0,t)+\frac{t}{12}} \right)\right]=\mathrm{Det}(I-\sigma_{z,t}K_{\rm Ai, \Gamma})_{\mathbb{L}^2(\mathbb{R})}\;.
\end{equation}
where $\sigma_{z,t}$ is the Fermi factor previously introduced in Eq.~\eqref{M sw} and $K_{\rm Ai, \Gamma}$ is the deformed Airy kernel which expression is 
\begin{equation} \label{KAiGamma}
  K_{\rm Ai, \Gamma}(u,u') = \int_{0}^{+\infty} \rmd r \, \Ai_\Gamma^\Gamma(r+u,t^{-\frac{1}{3}},w,w) \Ai_\Gamma^\Gamma(r+u',t^{-\frac{1}{3}},w,w)\;,
\end{equation}
where the deformed Airy function is equal to
\begin{equation} \label{aigamma} 
\Ai_\Gamma^\Gamma(a,b,c,d) := \int\limits_{\mathbb{R}+i\epsilon} \frac{\rmd\eta}{2 \pi} \mathrm{exp}\Big(i \frac{\eta^3}{3}+ i a\eta\Big)\frac{\Gamma(i b\eta+d)}{\Gamma(-i b\eta+c)}\;.
\end{equation}
where $\Gamma$ is the  Gamma function and $\epsilon \in [0, \Re(d/b))$ due to the pole of the $\Gamma$ function.


\section{The Painlevé II hierarchy}\label{app:PIIhierarchy}

We present in this Appendix a short review of the Painlevé II hierarchy. We will follow the conventions of Ref.~\cite[Eqs.~(1.31-1.32)]{claeys2010higher} which were also the ones of  \cite{doussal2018multicritical,cafasso2019fredholm}. The Painlevé II hierarchy is a sequence of ordinary non-linear differential equations obtained recursively upon the action of the Lenard operators $\{\mathcal{L}_k \}$ as 
\begin{equation}
\big(\frac{\rmd}{\rmd s}+2q\big)\mathcal{L}_n[q'-q^2]+\big(\frac{\rmd}{\rmd s}+2q\big)\sum_{k=1}^{n-1} \tau_k \mathcal{L}_k[q'-q^2]=sq.
\end{equation}
The operators $\mathcal{L}_k$ are defined recursively as 
\begin{equation}
\frac{\rmd }{\rmd s}\mathcal{L}_{j+1}f=\big( \frac{\rmd^3}{\rmd s^3}+4f \frac{\rmd}{\rmd s}+2f'\big)\mathcal{L}_j f,
\end{equation}
with the initial condition and the constraints
\begin{equation}
\mathcal{L}_0f=\frac{1}{2}, \qquad \mathcal{L}_j 1=0,\;  j\geq 1.
\end{equation}
In particular, we get for the first operators $\mathcal{L}_1f=f$, $\mathcal{L} _2f=f''+3f^2$ and the first equations of the hierarchy read
\begin{equation}
\begin{split}
&q''=2q^3+sq\\
&q''''= 10q (q')^2+10q^2q''-6q^5 -\tau_1(q''-2q^3)+sq
\end{split}
\end{equation}

The work of Refs.~\cite{doussal2018multicritical,cafasso2019fredholm}, generalizing the results of Hastings-McLeod \cite{hastings1980boundary} and Ablowitz-Segur \cite{segur1981asymptotic}, showed that the $n$-th member of the hierarchy can be solved for real $s$ by a real-valued function without pole upon a particular choice of asymptotic condition. Let $A$ be the solution of the differential equation 
\begin{equation}
\partial_x^{2n} A(x)+\sum_{k=1}^{n-1} \tau_k \partial_x^{2k} A(x)=xA(x).
\end{equation}
where all parameters are real-valued. Imposing the asymptotic condition $q(s)\underset{s\to +\infty}{\sim} \sqrt{\gamma}A(s)$, with $\gamma\in [0,1]$, the solution of the $n$-th member of the hierarchy reads with our notations 
\begin{equation}
q(s)=\bra{\delta}\frac{ \sqrt{\gamma} A_s}{I-\gamma K_s}\ket{\delta}=\frac{1}{2}\bra{\delta}\frac{ I}{I-\sqrt{\gamma} A_s}-\frac{ I}{I+\sqrt{\gamma} A_s}\ket{\delta}.
\end{equation}

Furthermore, as a byproduct, let us notice that these solutions of the hierarchy are numerically tractable as long as computing the function $A$ is. The reason for that is the identity
\begin{equation}
q(s)=\sqrt{-\frac{\rmd^2}{\rmd s^2}\log \Det[I-\gamma K_s]}
\end{equation}
and the fact that the Fredholm determinant is itself computable by Bornemann's method conditioned on an easy numerical evaluation of $K_s$ \cite{bornemann2010numerical,bornemann2011accuracy}.

\bibliographystyle{unsrt}
\bibliography{references}

\begin{thebibliography}{10}

\bibitem{babelon2003introduction}
Olivier Babelon, Denis Bernard, and Michel Talon.
\newblock {\em Introduction to classical integrable systems}.
\newblock Cambridge University Press, 2003.

\bibitem{faddeev2007hamiltonian}
Ludwig Faddeev and Leon Takhtajan.
\newblock {\em Hamiltonian methods in the theory of solitons}.
\newblock Springer Science \& Business Media, 2007.

\bibitem{gaudin2014bethe}
Michel Gaudin.
\newblock {\em The Bethe Wavefunction}.
\newblock Cambridge University Press, 2014.

\bibitem{soshnikov2000determinantal}
Alexander Soshnikov.
\newblock Determinantal random point fields.
\newblock {\em Russian Mathematical Surveys}, 55(5):923, 2000.

\bibitem{spohn2012large}
Herbert Spohn.
\newblock {\em Large scale dynamics of interacting particles}.
\newblock Springer Science \& Business Media, 2012.

\bibitem{mehta2004random}
Madan~Lal Mehta.
\newblock {\em Random matrices}.
\newblock Elsevier, 2004.

\bibitem{anderson2010introduction}
Greg~W Anderson, Alice Guionnet, and Ofer Zeitouni.
\newblock {\em An introduction to random matrices}, volume 118.
\newblock Cambridge university press, 2010.

\bibitem{potters2019first}
Marc Potters and Jean-Philippe Bouchaud.
\newblock A first course in random matrix theory.
\newblock 2019.

\bibitem{baik2016combinatorics}
Jinho Baik, Percy Deift, and Toufic Suidan.
\newblock {\em Combinatorics and random matrix theory}, volume 172.
\newblock American Mathematical Soc., 2016.

\bibitem{tracy1994level}
Craig~A Tracy and Harold Widom.
\newblock Level-spacing distributions and the {A}iry kernel.
\newblock {\em Communications in Mathematical Physics}, 159(1):151--174, 1994.

\bibitem{tracy1996orthogonal}
Craig~A Tracy and Harold Widom.
\newblock On orthogonal and symplectic matrix ensembles.
\newblock {\em Communications in Mathematical Physics}, 177(3):727--754, 1996.

\bibitem{borodin2002fredholm}
Alexei Borodin and Percy Deift.
\newblock Fredholm determinants, jimbo-miwa-ueno $\tau$-functions, and
  representation theory.
\newblock {\em Communications on Pure and Applied Mathematics: A Journal Issued
  by the Courant Institute of Mathematical Sciences}, 55(9):1160--1230, 2002.

\bibitem{jimbo1981monodromy}
Michio Jimbo, Tetsuji Miwa, and Kimio Ueno.
\newblock Monodromy preserving deformation of linear ordinary differential
  equations with rational coefficients: I. general theory and $\tau$-function.
\newblock {\em Physica D: Nonlinear Phenomena}, 2(2):306--352, 1981.

\bibitem{dyson1962brownian}
Freeman~J Dyson.
\newblock A {B}rownian-motion model for the eigenvalues of a random matrix.
\newblock {\em Journal of Mathematical Physics}, 3(6):1191--1198, 1962.

\bibitem{deift1999orthogonal}
Percy Deift.
\newblock {\em Orthogonal polynomials and random matrices: a Riemann-Hilbert
  approach}, volume~3.
\newblock American Mathematical Soc., 1999.

\bibitem{deift1997riemann}
Percy~A Deift, Alexander~R Its, and Xin Zhou.
\newblock A riemann-hilbert approach to asymptotic problems arising in the
  theory of random matrix models, and also in the theory of integrable
  statistical mechanics.
\newblock {\em Annals of mathematics}, 146(1):149--235, 1997.

\bibitem{dyson1976fredholm}
Freeman~J Dyson.
\newblock Fredholm determinants and inverse scattering problems.
\newblock {\em Communications in Mathematical Physics}, 47(2):171--183, 1976.

\bibitem{faddeev1976inverse}
LD~Faddeev.
\newblock Inverse problem of quantum scattering theory. ii.
\newblock {\em Journal of Soviet Mathematics}, 5(3):334--396, 1976.

\bibitem{dean2015finite}
David~S Dean, Pierre Le~Doussal, Satya~N Majumdar, and Gr{\'e}gory Schehr.
\newblock Finite-temperature free fermions and the {K}ardar-{P}arisi-{Z}hang
  equation at finite time.
\newblock {\em Physical review letters}, 114(11):110402, 2015.

\bibitem{dean2016noninteracting}
David~S Dean, Pierre Le~Doussal, Satya~N Majumdar, and Gr{\'e}gory Schehr.
\newblock Noninteracting fermions at finite temperature in a d-dimensional
  trap: Universal correlations.
\newblock {\em Physical Review A}, 94(6):063622, 2016.

\bibitem{le2018multicritical}
Pierre Le~Doussal, Satya~N Majumdar, and Gr{\'e}gory Schehr.
\newblock Multicritical edge statistics for the momenta of fermions in
  nonharmonic traps.
\newblock {\em Physical review letters}, 121(3):030603, 2018.

\bibitem{stephan2019free}
Jean-Marie St{\'e}phan.
\newblock Free fermions at the edge of interacting systems.
\newblock {\em SciPost Phys}, 6:057, 2019.

\bibitem{fokas1992isomonodromy}
Athanassios~S Fokas, AR~Its, and AV~Kitaev.
\newblock The isomonodromy approach to matric models in 2d quantum gravity.
\newblock {\em Communications in Mathematical Physics}, 147(2):395--430, 1992.

\bibitem{fokas1991discrete}
AS~Fokas, AR~Its, and AV~Kitaev.
\newblock Discrete painlev{\'e} equations and their appearance in quantum
  gravity.
\newblock {\em Communications in Mathematical Physics}, 142(2):313--344, 1991.

\bibitem{forrester2011non}
Peter~J Forrester, Satya~N Majumdar, and Gr{\'e}gory Schehr.
\newblock Non-intersecting brownian walkers and yang--mills theory on the
  sphere.
\newblock {\em Nuclear Physics B}, 844(3):500--526, 2011.

\bibitem{di19932d}
Philippe Di~Francesco, Paul Ginsparg, and Jean Zinn-Justin.
\newblock 2d gravity and random matrices.
\newblock {\em arXiv preprint hep-th/9306153}, 1993.

\bibitem{stanford2019jt}
Douglas Stanford and Edward Witten.
\newblock Jt gravity and the ensembles of random matrix theory.
\newblock {\em arXiv preprint arXiv:1907.03363}, 2019.

\bibitem{baik2000limiting}
Jinho Baik and Eric~M Rains.
\newblock Limiting distributions for a polynuclear growth model with external
  sources.
\newblock {\em Journal of Statistical Physics}, 100(3-4):523--541, 2000.

\bibitem{baik2018pfaffian}
Jinho Baik, Guillaume Barraquand, Ivan Corwin, Toufic Suidan, et~al.
\newblock Pfaffian schur processes and last passage percolation in a
  half-quadrant.
\newblock {\em The Annals of Probability}, 46(6):3015--3089, 2018.

\bibitem{imamura2004fluctuations}
Takashi Imamura and Tomohiro Sasamoto.
\newblock Fluctuations of the one-dimensional polynuclear growth model with
  external sources.
\newblock {\em Nuclear Physics B}, 699(3):503--544, 2004.

\bibitem{prahofer2000universal}
Michael Pr{\"a}hofer and Herbert Spohn.
\newblock Universal distributions for growth processes in 1+ 1 dimensions and
  random matrices.
\newblock {\em Physical review letters}, 84(21):4882, 2000.

\bibitem{sasamoto2010one}
Tomohiro Sasamoto and Herbert Spohn.
\newblock One-dimensional {K}ardar-{P}arisi-{Z}hang equation: an exact solution
  and its universality.
\newblock {\em Physical review letters}, 104(23):230602, 2010.

\bibitem{CalabreseDR}
P.~Calabrese, P.~Le~Doussal, and A.~Rosso.
\newblock Free-energy distribution of the directed polymer at high temperature.
\newblock {\em EPL (Europhysics Letters)}, 90(2):20002, 2010.

\bibitem{calabrese2011exact}
Pasquale Calabrese and Pierre Le~Doussal.
\newblock Exact solution for the {K}ardar-{P}arisi-{Z}hang equation with flat
  initial conditions.
\newblock {\em Physical review letters}, 106(25):250603, 2011.

\bibitem{ACQ}
G.~Amir, I.~Corwin, and J.~Quastel.
\newblock Probability distribution of the free energy of the continuum directed
  random polymer in 1 + 1 dimensions.
\newblock {\em Communications on Pure and Applied Mathematics}, 64:466--537,
  2011.

\bibitem{dotsenko}
V.~Dotsenko.
\newblock Bethe ansatz derivation of the {T}racy-{W}idom distribution for
  one-dimensional directed polymers.
\newblock {\em EPL (Europhysics Letters)}, 90(2):20003, 2010.

\bibitem{barraquand2020half}
Guillaume Barraquand, Alexandre Krajenbrink, and Pierre~Le Doussal.
\newblock Half-space stationary {K}ardar-{P}arisi-{Z}hang equation.
\newblock {\em arXiv:2003.03809}, 2020.

\bibitem{quastel2015one}
Jeremy Quastel and Herbert Spohn.
\newblock The one-dimensional kpz equation and its universality class.
\newblock {\em Journal of Statistical Physics}, 160(4):965--984, 2015.

\bibitem{novikov1984theory}
S~Novikov, SV~Manakov, LP~Pitaevskii, and Vladimir~E Zakharov.
\newblock {\em Theory of solitons: the inverse scattering method}.
\newblock Springer Science \& Business Media, 1984.

\bibitem{poppe1988fredholm}
Ch~P{\"o}ppe.
\newblock Fredholm determinants and the $\tau$ function for the
  kadomtsev-petviashvili hierarchy.
\newblock {\em Publications of the Research Institute for Mathematical
  Sciences}, 24(4):505--538, 1988.

\bibitem{poppe1984fredholm}
Christoph P{\"o}ppe.
\newblock The fredholm determinant method for the kdv equations.
\newblock {\em Physica D: Nonlinear Phenomena}, 13(1-2):137--160, 1984.

\bibitem{kulesza2012determinantal}
Alex Kulesza and Ben Taskar.
\newblock Determinantal point processes for machine learning.
\newblock {\em arXiv preprint arXiv:1207.6083}, 2012.

\bibitem{forrester2010log}
Peter~J Forrester.
\newblock {\em Log-gases and random matrices (LMS-34)}.
\newblock Princeton University Press, 2010.

\bibitem{andreief1883note}
C~Andr{\'e}ief.
\newblock Note sur une relation les int{\'e}grales d{\'e}finies des produits
  des fonctions.
\newblock {\em M{\'e}m. de la Soc. Sci. Bordeaux}, 2(1):1--14, 1883.

\bibitem{forrester2018meet}
Peter~J Forrester.
\newblock Meet {A}ndr{\'e}ief, {B}ordeaux 1886, and {A}ndreev, {K}harkov
  1882--1883.
\newblock {\em Random Matrices: Theory and Applications}, page 1930001, 2018.

\bibitem{baik2020largest}
Jinho Baik and Thomas Bothner.
\newblock The largest real eigenvalue in the real {G}inibre ensemble and its
  relation to the {Z}akharov--{S}habat system.
\newblock {\em The Annals of Applied Probability}, 30(1):460--501, 2020.

\bibitem{calabrese2004entanglement}
Pasquale Calabrese and John Cardy.
\newblock Entanglement entropy and quantum field theory.
\newblock {\em Journal of Statistical Mechanics: Theory and Experiment},
  2004(06):P06002, 2004.

\bibitem{calabrese2009entanglement}
Pasquale Calabrese and John Cardy.
\newblock Entanglement entropy and conformal field theory.
\newblock {\em Journal of Physics A: Mathematical and Theoretical},
  42(50):504005, 2009.

\bibitem{calabrese2015random}
Pasquale Calabrese, Pierre Le~Doussal, and Satya~N Majumdar.
\newblock Random matrices and entanglement entropy of trapped {F}ermi gases.
\newblock {\em Physical Review A}, 91(1):012303, 2015.

\bibitem{brezin1998level}
E~Br{\'e}zin and S~Hikami.
\newblock Level spacing of random matrices in an external source.
\newblock {\em Physical Review E}, 58(6):7176, 1998.

\bibitem{johansson2005random}
Kurt Johansson.
\newblock Random matrices and determinantal processes.
\newblock {\em arXiv preprint math-ph/0510038}, 2005.

\bibitem{doussal2018multicritical}
Pierre~Le Doussal, Satya~N Majumdar, and Gr{\'e}gory Schehr.
\newblock Multicritical edge statistics for the momenta of fermions in
  non-harmonic traps.
\newblock {\em arXiv preprint arXiv:1802.06436}, 2018.

\bibitem{ferrari2005determinantal}
Patrik~L Ferrari and Herbert Spohn.
\newblock A determinantal formula for the {GOE} {T}racy--{W}idom distribution.
\newblock {\em Journal of Physics A: Mathematical and General}, 38(33):L557,
  2005.

\bibitem{sasamoto2005spatial}
Tomohiro Sasamoto.
\newblock Spatial correlations of the {1D} {KPZ} surface on a flat substrate.
\newblock {\em Journal of Physics A: Mathematical and General}, 38(33):L549,
  2005.

\bibitem{gueudre2012directed}
Thomas Gueudr{\'e} and Pierre Le~Doussal.
\newblock Directed polymer near a hard wall and {KPZ} equation in the
  half-space.
\newblock {\em EPL (Europhysics Letters)}, 100(2):26006, 2012.

\bibitem{bohigas2009deformations}
O~Bohigas, JX~De~Carvalho, and Mauricio~Porto Pato.
\newblock Deformations of the {T}racy--{W}idom distribution.
\newblock {\em Physical review E}, 79(3):031117, 2009.

\bibitem{bothner2018large}
Thomas Bothner and Robert Buckingham.
\newblock Large deformations of the {T}racy--{W}idom distribution {I}:
  non-oscillatory asymptotics.
\newblock {\em Communications in Mathematical Physics}, 359(1):223--263, 2018.

\bibitem{rains2000correlation}
Eric~M Rains.
\newblock Correlation functions for symmetrized increasing subsequences.
\newblock {\em arXiv preprint math/0006097}, 2000.

\bibitem{ortmann2017pfaffian}
Janosch Ortmann, Jeremy Quastel, and Daniel Remenik.
\newblock A pfaffian representation for flat asep.
\newblock {\em Communications on Pure and Applied Mathematics}, 70(1):3--89,
  2017.

\bibitem{le2012kpz}
Pierre Le~Doussal and Pasquale Calabrese.
\newblock The kpz equation with flat initial condition and the directed polymer
  with one free end.
\newblock {\em Journal of Statistical Mechanics: Theory and Experiment},
  2012(06):P06001, 2012.

\bibitem{forrester2000painlev}
PJ~Forrester.
\newblock Painlev{\'e} transcendent evaluation of the scaled distribution of
  the smallest eigenvalue in the {L}aguerre orthogonal and symplectic
  ensembles.
\newblock {\em arXiv preprint nlin/0005064}, 2000.

\bibitem{a2005matrix}
Craig A~Tracy and Harold Widom.
\newblock Matrix kernels for the gaussian orthogonal and symplectic ensembles.
\newblock In {\em Annales de l'institut Fourier}, volume~55, pages 2197--2207,
  2005.

\bibitem{rumanov2016painleve}
Igor Rumanov.
\newblock Painlev{\'e} representation of {T}racy--{W}idom$_\beta$ distribution
  for $\beta=6$.
\newblock {\em Communications in Mathematical Physics}, 342(3):843--868, 2016.

\bibitem{grava2016tracy}
Tamara Grava, Alexander Its, Andrei Kapaev, and Francesco Mezzadri.
\newblock On the {T}racy-{W}idom $\beta$ distribution for $\beta= 6$.
\newblock {\em SIGMA. Symmetry, Integrability and Geometry: Methods and
  Applications}, 12:105, 2016.

\bibitem{quastel2019kp}
Jeremy Quastel and Daniel Remenik.
\newblock {KP} governs random growth off a one dimensional substrate.
\newblock {\em arXiv preprint arXiv:1908.10353}, 2019.

\bibitem{doussal2019large}
Pierre Le~Doussal.
\newblock Large deviations for the {Kardar-- Parisi--Zhang equation from the
  Kadomtsev--Petviashvili equation}.
\newblock {\em Journal of Statistical Mechanics: Theory and Experiment},
  2020(4):043201, 2020.

\bibitem{claeys2020forth}
Mattia Cafasso and Tom Claeys.
\newblock The {K}d{V} equation, multiplicative statistics for the {A}iry point
  process and the {KPZ} equation.
\newblock {\em In preparation}, 2020.

\bibitem{johansson2018gaussian}
Kurt Johansson, Gaultier Lambert, et~al.
\newblock Gaussian and non-gaussian fluctuations for mesoscopic linear
  statistics in determinantal processes.
\newblock {\em The Annals of Probability}, 46(3):1201--1278, 2018.

\bibitem{majumdar2011many}
Satya~N Majumdar, C{\'e}line Nadal, Antonello Scardicchio, and Pierpaolo Vivo.
\newblock How many eigenvalues of a gaussian random matrix are positive?
\newblock {\em Physical Review E}, 83(4):041105, 2011.

\bibitem{majumdar2012number}
Satya~N Majumdar and Pierpaolo Vivo.
\newblock Number of relevant directions in principal component analysis and
  wishart random matrices.
\newblock {\em Physical review letters}, 108(20):200601, 2012.

\bibitem{krajenbrink2018systematic}
Alexandre Krajenbrink, Pierre Le~Doussal, and Sylvain Prolhac.
\newblock Systematic time expansion for the kardar--parisi--zhang equation,
  linear statistics of the gue at the edge and trapped fermions.
\newblock {\em Nuclear Physics B}, 936:239--305, 2018.

\bibitem{krajenbrink2019linear}
Alexandre Krajenbrink and Pierre Le~Doussal.
\newblock Linear statistics and pushed coulomb gas at the edge of
  $\beta$-random matrices: Four paths to large deviations.
\newblock {\em EPL (Europhysics Letters)}, 125(2):20009, 2019.

\bibitem{grabsch2017truncated}
Aur{\'e}lien Grabsch, Satya~N Majumdar, and Christophe Texier.
\newblock Truncated linear statistics associated with the top eigenvalues of
  random matrices.
\newblock {\em Journal of Statistical Physics}, 167(2):234--259, 2017.

\bibitem{krajenbrink2019beyond}
Alexandre Krajenbrink.
\newblock {\em Beyond the typical fluctuations: a journey to the large
  deviations in the Kardar-Parisi-Zhang growth model}.
\newblock PhD thesis, PSL Research University, 2019.

\bibitem{bothner2020origins}
Thomas Bothner.
\newblock On the origins of {R}iemann-{H}ilbert problems in mathematics.
\newblock {\em arXiv preprint arXiv:2003.14374}, 2020.

\bibitem{johansson2017edge}
Kurt Johansson.
\newblock Edge fluctuations of limit shapes.
\newblock {\em arXiv preprint arXiv:1704.06035}, 2017.

\bibitem{cafasso2019fredholm}
Mattia Cafasso, Tom Claeys, and Manuela Girotti.
\newblock Fredholm determinant solutions of the {P}ainlev{\'e} {II} hierarchy
  and gap probabilities of determinantal point processes.
\newblock {\em arXiv preprint arXiv:1902.05595}, 2019.

\bibitem{brezin2016random}
Edouard Br{\'e}zin and Shinobu Hikami.
\newblock {\em Random matrix theory with an external source}.
\newblock Springer, 2016.

\bibitem{balakrishnan1982inhomogeneous}
Radha Balakrishnan.
\newblock On the inhomogeneous {H}eisenberg chain.
\newblock {\em Journal of Physics C: Solid State Physics}, 15(36):L1305, 1982.

\bibitem{ablowitz1991solitons}
Mark~J Ablowitz, MA~Ablowitz, PA~Clarkson, and Peter~A Clarkson.
\newblock {\em Solitons, nonlinear evolution equations and inverse scattering},
  volume 149.
\newblock Cambridge university press, 1991.

\bibitem{shabat1972exact}
A~Shabat and V~Zakharov.
\newblock Exact theory of two-dimensional self-focusing and one-dimensional
  self-modulation of waves in nonlinear media.
\newblock {\em Soviet physics JETP}, 34(1):62, 1972.

\bibitem{rider2014extremal}
Brian Rider and Christopher~D Sinclair.
\newblock Extremal laws for the real {G}inibre ensemble.
\newblock {\em The Annals of Applied Probability}, 24(4):1621--1651, 2014.

\bibitem{poplavskyi2017distribution}
Mihail Poplavskyi, Roger Tribe, and Oleg Zaboronski.
\newblock On the distribution of the largest real eigenvalue for the real
  {G}inibre ensemble.
\newblock {\em The Annals of Applied Probability}, 27(3):1395--1413, 2017.

\bibitem{bornemann2010numerical}
Folkmar Bornemann.
\newblock On the numerical evaluation of {F}redholm determinants.
\newblock {\em Mathematics of Computation}, 79(270):871--915, 2010.

\bibitem{beals1984scattering}
Richard Beals and Ronald~R Coifman.
\newblock Scattering and inverse scattering for first order systems.
\newblock {\em Communications on Pure and Applied Mathematics}, 37(1):39--90,
  1984.

\bibitem{lakshmanan1977continuum}
M~Lakshmanan.
\newblock Continuum spin system as an exactly solvable dynamical system.
\newblock {\em Physics Letters A}, 61(1):53--54, 1977.

\bibitem{calabrese2010universal}
Pasquale Calabrese and Fabian~HL Essler.
\newblock Universal corrections to scaling for block entanglement in spin-1/2
  {XX} chains.
\newblock {\em Journal of Statistical Mechanics: Theory and Experiment},
  2010(08):P08029, 2010.

\bibitem{tracy1998correlation}
Craig~A Tracy and Harold Widom.
\newblock Correlation functions, cluster functions, and spacing distributions
  for random matrices.
\newblock {\em Journal of statistical physics}, 92(5-6):809--835, 1998.

\bibitem{krajenbrink2018large}
Alexandre Krajenbrink and Pierre~Le Doussal.
\newblock Large fluctuations of the {KPZ} equation in a half-space.
\newblock {\em SciPost Phys}, 5:032, 2018.

\bibitem{baik2008asymptotics}
Jinho Baik, Robert Buckingham, and Jeffery DiFranco.
\newblock Asymptotics of {T}racy--{W}idom distributions and the total integral
  of a painlev{\'e} ii function.
\newblock {\em Communications in Mathematical Physics}, 280(2):463--497, 2008.

\bibitem{bornemann2011accuracy}
Folkmar Bornemann.
\newblock Accuracy and stability of computing high-order derivatives of
  analytic functions by {C}auchy integrals.
\newblock {\em Foundations of Computational Mathematics}, 11(1):1--63, 2011.

\bibitem{bender2010edge}
Martin Bender.
\newblock Edge scaling limits for a family of non-{H}ermitian random matrix
  ensembles.
\newblock {\em Probability theory and related fields}, 147(1-2):241--271, 2010.

\bibitem{akemann2012universality}
Gernot Akemann and Michael~J Phillips.
\newblock Universality conjecture for all {A}iry, sine and {B}essel kernels in
  the complex plane.
\newblock {\em arXiv preprint arXiv:1204.2740}, 2012.

\bibitem{KPZ}
M.~Kardar, G.~Parisi, and Y.Z. Zhang.
\newblock Dynamic scaling of growing interfaces.
\newblock {\em Phys. Rev. Lett.}, 56:889--892, 1986.

\bibitem{imamura2012exact}
Takashi Imamura and Tomohiro Sasamoto.
\newblock Exact solution for the stationary {K}ardar-{P}arisi-{Z}hang equation.
\newblock {\em Physical review letters}, 108(19):190603, 2012.

\bibitem{borodin2015height}
Alexei Borodin, Ivan Corwin, Patrik Ferrari, and B{\'a}lint Vet{\H{o}}.
\newblock Height fluctuations for the stationary {KPZ} equation.
\newblock {\em Mathematical Physics, Analysis and Geometry}, 18(1):20, 2015.

\bibitem{claeys2010higher}
Tom Claeys, Igor Krasovsky, and Alexander Its.
\newblock Higher-order analogues of the {T}racy-{W}idom distribution and the
  {P}ainlev{\'e} {II} hierarchy.
\newblock {\em Communications on pure and applied mathematics}, 63(3):362--412,
  2010.

\bibitem{hastings1980boundary}
Stuart~P Hastings and John~Bryce Mcleod.
\newblock A boundary value problem associated with the second {P}ainlev{\'e}
  transcendent and the {K}orteweg-de {V}ries equation.
\newblock {\em Archive for Rational Mechanics and Analysis}, 73(1):31--51,
  1980.

\bibitem{segur1981asymptotic}
Harvey Segur and Mark~J Ablowitz.
\newblock Asymptotic solutions of nonlinear evolution equations and a painleve
  transcedent.
\newblock {\em Physica D: Nonlinear Phenomena}, 3(1-2):165--184, 1981.

\end{thebibliography}

\end{document}